\documentclass[]{article}
\usepackage{graphicx}
\usepackage{xcolor}
\usepackage{amsfonts}
\usepackage{amssymb}
\def\ligne#1{\hbox to\hsize{#1}}
\def\leurre{\noindent\leftskip0pt\small\baselineskip 10pt}
\newtheorem{thm}{\textrm{\sc Theorem}}

\newtheorem{fig}{\textrm{Figure}}
\newtheorem{tab}{\textrm{Table}}
\def\boxempty{\hbox{\vbox{\hsize=7pt\offinterlineskip
\ligne{
\vrule height 7pt depth 0pt width 0.6pt
\vbox to 7pt{\hsize=5.8pt
\hrule height 0pt depth 0.6pt width 5.8pt
\vfill
\hrule height 0.6pt depth 0pt width 5.8pt
}\hskip-0.5pt
\vrule height 7pt depth 0pt width 0.6pt
}}
}}
\def\trep{\hrule height 1pt depth 1pt width \hsize}
\def\trfn{\hrule height 0.5pt depth 0.5pt width \hsize}
\newcounter{laform}
\author{Maurice {\sc Margenstern}}
\title{A weakly universal cellular automaton in the heptagrid.}
\begin{document}
\maketitle

\begin{abstract}
In this paper, we construct a weakly universal cellular automaton in the heptagrid, 
the tessellation
$\{7,3\}$ which is not rotation invariant but which is truly planar. This
result, under these conditions, cannot be improved for the tessellations $\{p,3\}$.
\end{abstract}

\section{Introduction}
\def\zz#1{{\footnotesize\tt #1}}

   This paper is basically an improvement of papers~\cite{mmarXiv1605b} and
~\cite{mmarXiv1606} where I proved the
same result in the tessellations $\{9,3\}$ and $\{8,3\}$ respectively. The reason of 
this improvement lies in
the relatively small number of rules for paper~\cite{mmarXiv1605b} and the fact, noticed
in that paper, that several rules where uselessly duplicated. Also, as it is usual in 
this process of reducing the possibilities of the automaton, here its neighbourhood,
it is needed to change something in the previous scenario of the simulation.
Here, I repeat the scheme explained in~\cite{mmarXiv1606}. The key idea of that paper
was to combine two existing structures in order to eliminate one of the structures
used so far in the simulation scheme explained in~\cite{mmbook3}. Note that the
present result cannot be improved for this class of cellular automata which
explicitly make use of non rotation invariant rules: indeed, the heptagrid is
the tessellation $\{p,3\}$ of the hyperbolic plane with the smallest possible
value for~$p$ which is~7. In this paper, I use the new system of coordinates 
introduced in~\cite{mmarXiv1605a} for the tilings $\{p,3\}$ and $\{p$$-$$2,4\}$. 

   In this paper, the same model as in~\cite{mmJCA2016} and~\cite{mmarXiv1512} and the
other quoted papers is used. 

In Section~\ref{scenar}, I just indicate how the model is implemented in the
heptagrid. In Section~\ref{rules}, I give the
rules of the automaton, insisting in the way we defined these rules in a context where
rotation invariance is no more required, which allows us to prove the following result:

\begin{thm}\label{letheo}
There is a weakly universal cellular automaton on the heptagrid, the tessellation 
$\{7,3\}$
which is truly planar and which has two states.
\end{thm}

Presently, we turn to the proof of this result, repeating that the rules are not rotation
invariant: the statement of the theorem does not mention that condition.

\section{The scenario of the simulation}
\label{scenar}

   In the present section , I implement the standard structures used 
in~\cite{mmarXiv1606}: tracks, the passive fixed switch, the fork, the
doubler, the selector between a simple and a double locomotive, see further,
the controller and the sensor.

    I sketchily remember that we simulate a register machine by a railway circuit.
Such circuit assembles infinitely many portions of straight lines, quarters of circles 
and switches. There are three kinds of switches, see~\cite{stewart,mmbook3}, for
a description of the circuit and of its working. In~\cite{mmbook3} the simulation
is thoroughly described and it is adapted to the hyperbolic context.

    As in previous papers, the flip-flop and the memory switch are decomposed into
simpler ingredients which we call sensors and control devices. This reinforces the
importance of the tracks as their role for conveying key information is more and more 
decisive. Here too, tracks are blank cells marked by appropriate black cells we call
\textbf{milestones}. We carefully study this point in Sub-section~\ref{tracks}.  Later, 
in Sub-section~\ref{struct}, we adapt the configurations described in~\cite{mmarXiv1605b}
to the tessellation $\{8,3\}$.

\subsection{The tracks}\label{tracks}

In this implementation, the tracks are represented in a 
way which is a bit similar to that of~\cite{mmarXiv1605b,mmarXiv1606}. 
The present implementation is given by Figure~\ref{elemtrack}. 

Here, we explicitly indicate the numbering of
the sides in a cell which will be systematically used through the paper. As rotation
invariance is no more required, we fix a 
side which will be, by definition, side~1 in the considered cell. This choice
allows us to consider that the tracks are one way. The orientation is given
by the side~1 of each cell which is, by convention, the side shared by the next cell
on the track. All the other sides are numbered starting from this one and growing 
one by one while counter-clockwise turning around the cell. Note that, in our setting, 
the same side, which is shared by two cells, can receive two different numbers in the 
cells which share it. An example of 
this situation is given in Figure~\ref{elemtrack}: in the central cell we denote
by~0(0), side~1 is side~6 
in the neighbour of the central cell sharing this side. In Sub-section~\ref{srtrack}, 
we go back to the construction of the tracks starting from the elements indicated
in Figure~\ref{elemtrack}. Note that Figure~\ref{elemtrack} shows us two rays starting
from~$M$, the mid-point of the side~2 of the central cell. These rays allow us to
introduce the numbering of the tiles based on~\cite{mmarXiv1605a}. It will be used
in the figures illustrating the paper.

The rays delimit what we call a \textbf{sector}. The rays are defined as follows.
The ray~$u$ starts from the mid-point~$M$ of the side~2 of~0(0) and it
passes through the mid-point of its side~1. The ray also passes through the mid-point
of the side~7 of the neighbour of~0(0) which is seen through its side~1 and which we
denote 1(1). 
The ray~$v$, also issued from~$M$, cuts the sides~5 and~4 of~1(1) at their mid-points. 
Its support also passes through the mid-point of the side~3 of~0(0).

\vskip 10pt
\vtop{
\ligne{\hfill
\includegraphics[scale=1]{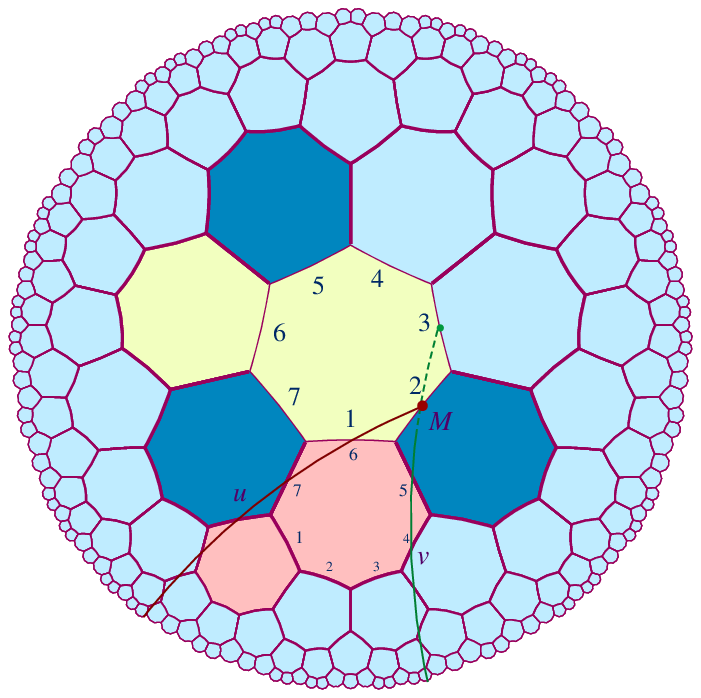}
\hfill}
\vspace{-10pt}
\ligne{\hfill
\vtop{\leftskip 0pt\parindent 0pt\hsize=200pt
\begin{fig}\label{elemtrack}
\leurre
Element of the tracks: in pink, the exit cells; in yellow: the entrance to the cell 
and the cell itself.
\end{fig}
}
\hfill}
}
\vskip 5pt
In the figures of the paper, the central cell
is the tile whose centre is the centre of the circle in which the figure is inscribed.
The central cell is numbered by~0, denoted by~0(0). We number the sides of the tile
as indicated in Figure~\ref{elemtrack}. For $i\in\{1..7\}$, the cell which
shares the side~$i$ with the central cell is called \textbf{neighbour}~$i$ and it
is denoted by~$1(i)$. The rotation around~0(0) allows us to attach a sector to each
tile~$1(i)$. Number~1 in this notation is the number given to the root of the
tree attached to the sector defined from this tile, see~\cite{mmbook2,mmarXiv1605a}.
Here, that definition is adapted to the case of the 
tessellation~$\{7,3\}$ which we call the heptagrid from now on as in many
previous papers. We invite the
reader to follow the present explanation on Figure~1. 
Consider the sector defined by the rays~$u$ and~$v$. The neighbours of the 
cell~1(1) sharing its sides~$j$, $j\in\{1..3\}$ are numbered~$j$+1 and are denoted
by $j$+1(1). We say that the cell~1(1) is a $W$-cell and its sons are defined
by the rule \hbox{\tt W $\rightarrow$ BWW}, which means that 2(1) is a $B$-cell.
This means that the sons of~2(1) are defined by the rule \hbox{\tt B $\rightarrow$ BW},
where the $B$-son has two consecutive sides crossed by~$u$ in their mid-points. 
These sons of~1(1) constitute the level~1 of the tree.
The sons of~2(1), starting from its $B$-son are numbered 5 and~6 denoted by
5(1) and 6(1) respectively. By induction, the level~$n$+1 of the tree
are the sons of the cells which lie on the level~$n$, applying the above rules. 
The cells are numbered from
the level~0, the root, level by level and, on each level from left to right, {\it i.e.}
starting from the ray~$u$ until the ray~$v$. What we have seen on the numbering of the
sons of~2(1) is enough to see how the process operates on the cells. From now on, we
use this numbering of the cells in the figures of the paper.

Further, Figure~\ref{stab_tracks} illustrates how to assemble elements of the
track on which the locomotive passes.
As mentioned in the caption, the trajectory of the locomotive is illustrated in yellow.
From the point of view of the cellular automaton, this is not a new state: yellow cells
are blank cells. This 
representation is used to facilitate the understanding by the reader. 
\vskip 10pt
\vtop{
\ligne{\hfill
\includegraphics[scale=1]{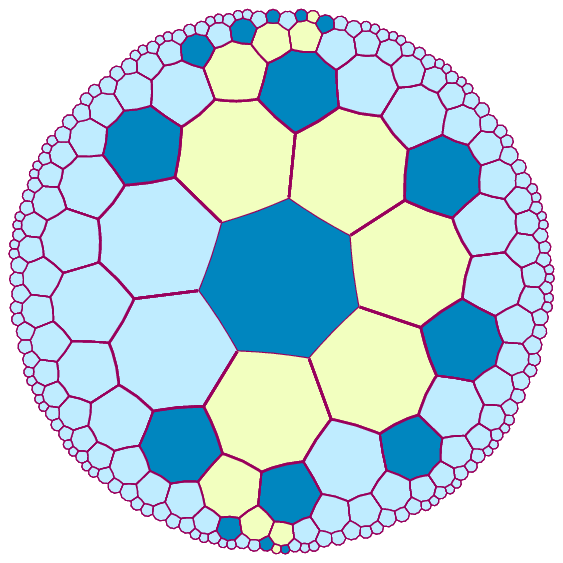}
\hfill}
\vspace{-10pt}
\ligne{\hfill
\vtop{\leftskip 0pt\parindent 0pt\hsize=200pt
\begin{fig}\label{stab_tracks}
\leurre
Element of the tracks: in yellow, the elements of the track where the locomotive
passes.
\end{fig}
}
\hfill}
}

As can be seen in Figures of Section~\ref{rules}, the locomotive is implemented as
a single black cell: it has the same colour as the milestones of the tracks. Only the 
position of the locomotive with respect to the milestones allows us to distinguish it 
from the milestones. As clear from the next sub-section, we know that besides this
\textbf{simple locomotive}, the locomotive also occurs as a \textbf{double one} in some
portions of the circuit: two consecutive black cells. In a double locomotive, we call 
its first, second cell the 
\textbf{front}, \textbf{rear} respectively of the locomotive. In a simple locomotive,
front and rear are the same cell which will be called front in that case. We reserve
the word rear for the second cell of a double locomotive.

We can see that in that figure that, assuming that the locomotive goes from top to
bottom, the milestones are in neighbours~2, 5 and~7 as for the cell 1(6) or
in neighbours~2, 4 and~6 as for the cell 3(1). It is important to notice that such 
tracks allow us to join any pair of points. In 
Section~\ref{rules}, we shall check that the rules will satisfy this constraint.

The circuit also makes use of signals which are implemented in the form of a simple 
locomotive. So that at some point, it may happen that we have three
simple locomotives travelling on the circuit: the locomotive and two auxiliary signals 
involved in the working of some switch. For aesthetic reasons, the black colour which 
is opposed to the blank is dark blue in the figures.

\subsection{The structures of the simulation}
\label{struct}

    The crossings of~\cite{stewart} are present in many ones of my papers. Starting 
from~\cite{mmarXiv1202}, I replaced the crossing by round-abouts, a road traffic
structure, in my simulations in the hyperbolic plane. At a round-about where two 
roads are crossing, if you want to keep the direction arriving at the round-about, 
you need to leave the round-about at the second road. I refer the reader 
to~\cite{mmbook3} for references. The structure is a complex one, which requires
a fixed switch, a doubler and a selector. Other structures are used to simulate the
switches used in~\cite{stewart,mmbook3}: the fork, the controller and the sensor.
In this section, we present the implementation of these structures which are those
of~\cite{mmarXiv1606} adapted to the present tessellation.

\subsubsection{The fixed switch, the doubler and the fork}
\label{ssfx}

   We look at the fixed switch first, and then at the doubler and the fork as the
doubler is a combination of the fork and of the fixed switch.

\vskip 10pt
\noindent
\textbf{The fixed switch}
\vskip 5pt
    As the tracks are one-way and as an active fixed switch always sends the locomotive 
in the same direction, no track is needed for the other direction: there is no active 
fixed switch. Now, passive fixed switches are still needed as just seen in the previous 
paragraph.  

\vtop{
\ligne{\hfill
\includegraphics[scale=1]{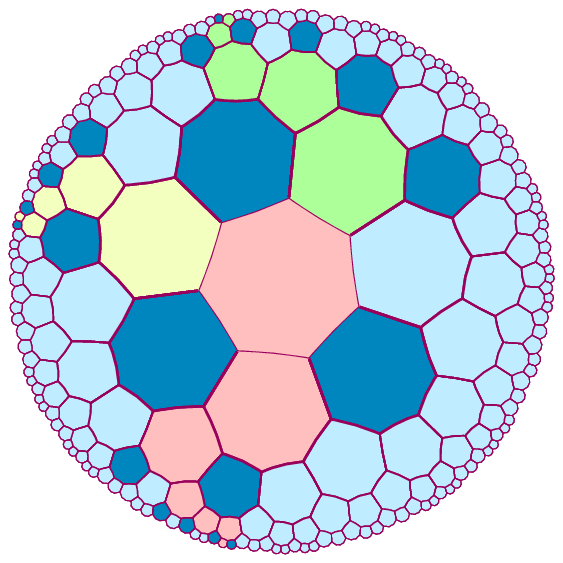}
\hfill}
\vspace{-20pt}
\ligne{\hfill
\vtop{\leftskip 0pt\parindent 0pt\hsize=270pt
\begin{fig}\label{stab_fx}\leurre
Idle configuration of the passive fixed switch. In yellow, the arriving path from
the left, in pink, the arriving path from the right, in pink, the path leaving the
switch.
\end{fig}
}
\hfill}
\vskip 10pt
}

Figure~\ref{stab_fx} illustrates the passive fixed switch when there is no 
locomotive around: we say that such a configuration is \textbf{idle}. We shall
again use this term in the similar situation for the other structures and for
individual cells too. 

We can see that it consists of elements of the tracks which are 
simply assembled in the appropriate way in order to drive the locomotive to the bottom 
direction in the picture, whatever upper side the locomotive arrived at the switch. 
The path followed by the locomotive to the switch is in yellow or in green until the 
central cell which is pink. The path from the left-hand side, yellow in the figure, 
consists, in this order of the cells 29(2), 11(2), 10(2), 3(2) and  1(2). From the 
right-hand side, green in the figure, it consists of the cells 23(1), 9(1), 3(1), 2(1) 
and~1(7). Of course, 1(2) and~1(7) are neighbours of~0(0). The path followed by the 
locomotive from 0(0) is in pink in the figure.
It consists of the following cells in this order: 0(0), 1(4), 2(4), 7(4), 8(4),
9(4) and 24(4). Note that the cell~0(0) in Figure~\ref{stab_fx} is a standard element
of the track with three milestones in 1(5), 1(1) and~1(3), its neighbours~2, 5 and~7
respectively.
Note that 1(3) and 1(1) are milestones for the cell 1(2), that 1(3) and~1(5) are 
milestones for 1(4) and that 2(7) and~1(1) are milestones for 1(7). Note that
the milestones of~1(7) are its neighbours~3, 5 and~7

From our description of the working of the round-about, a passive fixed switch must be 
crossed by a double locomotive as well as a simple one. Later, in 
Subsection~\ref{srfx}, we shall check that the structure illustrated by
Figure~\ref{stab_fx} allows those crossings. 

\vskip 10pt
\noindent
\textbf{The doubler and the fork}
\label{sssdblfrk}
\vskip 5pt
    The fork is the structure illustrated by the left-hand side picture of
 Figure~\ref{stab_dblfrk}. Note that its structure is very different 
from that of the tracks or of the fixed switch. The central cell~0(0) is 
black and two paths start from~1(1), each one on one side of the central cell
with respect to its axis which crosses its side~1 and passes through the vertex
which is opposite to side~1, it is shared by sides~4 and~5. The paths take each one two
cells around~0(0) and then leave the neighbourhood of the cell~0(0).
The cells leading to~1(1) are yellow in the figure. The left-hand side track is 
green, consisting of the following cells, in this order: 1(2), 1(3), 3(3), 7(3) 
and~20(3). The right-hand side track is pink. It consists of the cells: 1(7), 1(6), 4(6),
5(7) and 13(7). The locomotive, a simple one, arrives through the yellow path: 
28(1), 10(1), 4(1), and 1(1). From 1(1), two simple locomotives appear: one in 1(2), going 
on along the green path, 
the other in 1(7), travelling along the pink path. 

   The doubler is a structure which receives a simple locomotive and which yields a
double one. The idea is to use a fork to produce two simple locomotives and then
to gather them at a fixed switch in order to produce the double locomotive. The process 
is illustrated by the right-hand side of Figure~\ref{stab_dblfrk}.
The structure is inspired by that of~\cite{mmarXiv1606} but it turns out that here, it
is much simpler than there. The reason is that in~\cite{mmarXiv1606}, the even number of
sides compelled me to devise a detour in order that two locomotives arrive at the same time
one after the other at one entrance of the fixed switch. In the heptagrid, the odd number
of sides allowed me to perform a simpler implementation. The odd number allows us to 
have two equal paths around the common milestones of the concerned elements of the 
tracks. It is enough to place the central cell of the fixed switch at the end of one of 
the paths, the green path on the right-hand side picture of Figure~\ref{stab_dblfrk}.
The picture uses the same colours as the picture of
the fork with the same meaning. Consider the green path. Its cells are, in this
order: 2(2), 1(2)and 0(0) which makes three cells. The pink path
consists of the following cells, in this order: 3(1), 2(1) and 1(7).
We can see that the cells around 0(0) are exactly the neighbours of the central cell
of a fixed switch, see Figure~\ref{stab_fx}. According to this description,
the two simple locomotives created at the same time in 2(2) and 3(1) respectively
do not arrive at the same time at the cell~0(0). When the locomotive created in~2(2)
arrives at the cell~0(0), the locomotive created in 3(1) is at 1(7), so that
the two black cells in 1(7) and~0(0) constitute a double locomotive arriving from
the right whose front is in the central cell of the fixed switch. Then the double
locomotive leaves the switch through the orange path~: in this order, 1(4), 2(4),
7(4), 8(4), 9(4) and 24(4). Accordingly the structure works as expected for a doubler. 
Note that elements of the track only are involved.

\vskip 10pt
\vtop{
\ligne{\hfill
\includegraphics[scale=0.7]{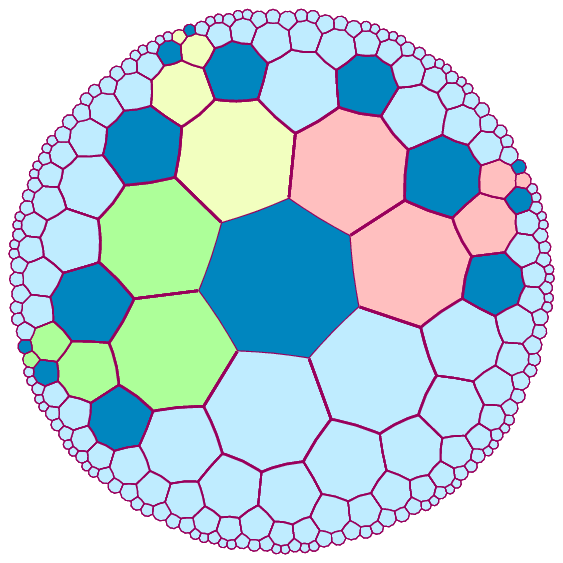}
\includegraphics[scale=0.7]{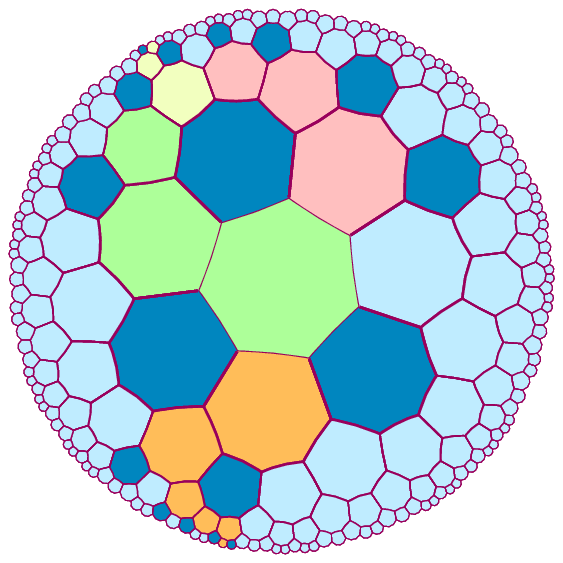}
\hfill}
\vspace{-15pt}
\ligne{\hfill
\vtop{\leftskip 0pt\parindent 0pt\hsize=300pt
\begin{fig}\label{stab_dblfrk}
\leurre
Idle configurations.
To left: the fork. To right: the doubler. In both of them: arrival of the locomotive
through the yellow track. Then, one locomotive on the green and on the pink tracks.
In the doubler: the double locomotive leaves the switch through the orange track.
\end{fig}
}
\hfill}
}

\subsubsection{The selector}
\label{ssel}

The selector is illustrated by Figure~\ref{stab_sel}. This structure is less symmetric
than the corresponding structure of~\cite{mmarXiv1606}, which makes another difference
with that paper. We have a yellow track through which the locomotive arrives, simple
or double, both cases are possible. When a simple locomotive arrives, it leaves the
cell through~1(1), via the pink path which consists of the cells 1(1), 2(1), 7(1)
and 18(1). When a double locomotive arrives, a simple locomotive leaves the structure
through the green path, the cells: 1(4), 2(5), 5(5), 12(4) and 33(4). Both cells 
1(5) and 1(7) can detect whether the locomotive is simple or double. They can do that 
when the front of the locomotive is in~0(0). Then, if the locomotive is double, its rear 
is in 1(6). Both cells 0(0) and~1(6) are neighbours of 1(5) and of 1(7) too.

In Subsubsection~\ref{ssel}, the rules will show that such a working will be observed.

\vtop{
\ligne{\hfill
\includegraphics[scale=1]{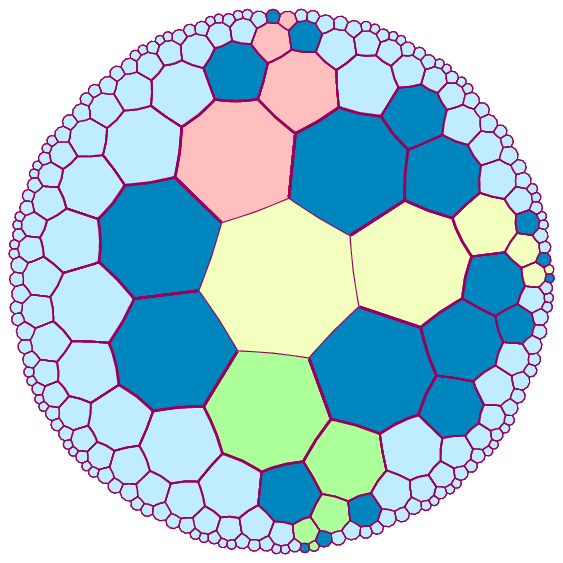}
\hfill}
\vspace{-15pt}
\ligne{\hfill
\vtop{\leftskip 0pt\parindent 0pt\hsize=300pt
\begin{fig}\label{stab_sel}
\leurre
Idle configuration of the selector. The cells~$1(7)$ and$~1(5)$ detect whether the 
locomotive is simple or double. Arriving through the yellow track, a simple locomotive
leaves through the pink track, a double locomotive leaves through the green track as
a simple locomotive.
\end{fig}
}
\hfill}
}

\subsection{The controller and the sensor}
\label{sctrlcapt}

    In this Sub-section, we look at the additional structures used for the
flip-flop and for the memory switch, see~\cite{stewart,mmbook3} for the definitions
and for the implementation in the hyperbolic plane. As explained in~\cite{mmbook3},
the flip-flop and the active memory switch are implemented by using the fixed switch,
the fork and a new structure we shall study in Subsubsection~\ref{ssctrl}: 
the \textbf{controller}. The structure is illustrated by Figure~\ref{stab_ctrl}.
For the passive memory switch, we need the fork, the fixed switch and another
new structure we shall study in Subsubsection~\ref{sscapt}: the \textbf{sensor}
illustrated by Figure~\ref{stab_capt}.

\subsubsection{The controller}\label{ssctrl}

As shown by Figure~\ref{stab_ctrl}, the controller sits on an ordinary cell
of the track. The locomotive which runs on that track is always simple.
The track consists of the yellow path which passes through 25(6), 9(6), 10(6), 4(6) 
and~1(6), a neihgbour of~0(0), and of the pink path which
starts from~0(0) and which is crossed by the locomotive when the controller is black.
The \textbf{colour} of the controller is defined by the cell~1(3), in orange in
Figure~\ref{stab_ctrl}. The pink path consists of the following cells, in this order:
0(0), 1(4), 2(5), 5(5), 12(4) and~33(4), a path already seen in previous figures.
When the cell~1(3) is black, then the cell~0(0) is an ordinary element of the track, 
so that the locomotive goes on its way along the pink path, leaving the controller. If 
the cell 1(3) is white, then the cell 0(0) can no more work as an element of the track. 
It remains white, which means that the locomotive is stopped at 1(6): after that, it 
vanishes. This corresponds to the working of a selection in an active passage of the 
switch: the locomotive cannot run along a non-selected track. Here it can do it for a 
while, but at some point, it is stopped by the controller. Note that the occurrence of a 
locomotive in the structure does not change the colour in 1(3). The change of colour 
in that cell is performed by a signal which takes the view of a simple locomotive arriving through another track: 
31(3), 12(3) and 4(3), that latter cell being a neighbour of~1(3). When the
locomotive-signal arrives at 4(3), it makes the cell 1(3) change its colour: from
white to black and from black to white.

\vtop{
\vskip 0pt
\ligne{\hfill
\includegraphics[scale=1]{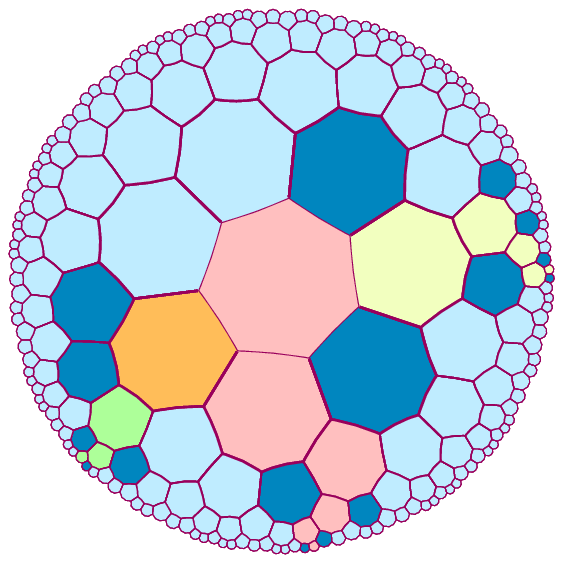}
\hfill}
\vspace{-20pt}
\ligne{\hfill
\vtop{\leftskip 0pt\parindent 0pt\hsize=320pt
\begin{fig}\label{stab_ctrl}
\leurre
Idle configuration of the controller of the flip-flop and of the active memory switch.
In orange, the cell~$1(3)$: the sensor which controls the working of the device.
In pink, the portion of the track which is allowed when the cell~$1(3)$ is black only.
\end{fig}
}
\hfill}
\vskip 10pt
}

\subsubsection{The sensor}\label{sscapt}

Let us now turn to the sensor,
illustrated by Figure~\ref{stab_capt}. As suggested by its name, the sensor does not
exactly behave like the controller. When the locomotive passes on the non-selected track,
it is not stopped. A fork creates two simple locomotives, 
see~\cite{mmarXiv1605b,mmarXiv1606}: one of them goes on to the 
switch, the other to the sensor which uses it as a messenger for the signal it has 
to send to the active switch in order to change the selection of the tracks. 

This is illustrated by the structure of the figure. The path is the same as in 
Figure~\ref{stab_ctrl}. The cell which plays the role of a sensor is this time the
cell~1(1) whose state we call the \textbf{colour} of the sensor. Note that the 
neighbourhood of that cell in Figure~\ref{stab_capt} is the 
same, up to rotation, to the neighbourhood of the cell~1(3) in Figure~\ref{stab_ctrl}:
the green path here consists of the cells 26(1), 9(1) and 3(1) the latter being a 
neighbour of~1(1).

Figure~\ref{stab_capt} shows a very different structure for the cell~0(0) compared
with that of Figure~\ref{stab_ctrl}. When the sensor is white, its neighbourhood
is exactly that of the cell~0(0) when the controller is black: it is an ordinary element
of the track so that the locomotive goes on its way on the track. The difference
in both structures lies in the logic of the switches. In the case of the controller,
when the locomotive goes on its way, it is the locomotive of the circuit going to
another switch or to a round-about. In the case of the sensor, the locomotive which
goes on its way on the track becomes a signal sent to the active switch associated
to the passive switch.

\vskip 10pt
\vtop{
\ligne{\hfill
\includegraphics[scale=1]{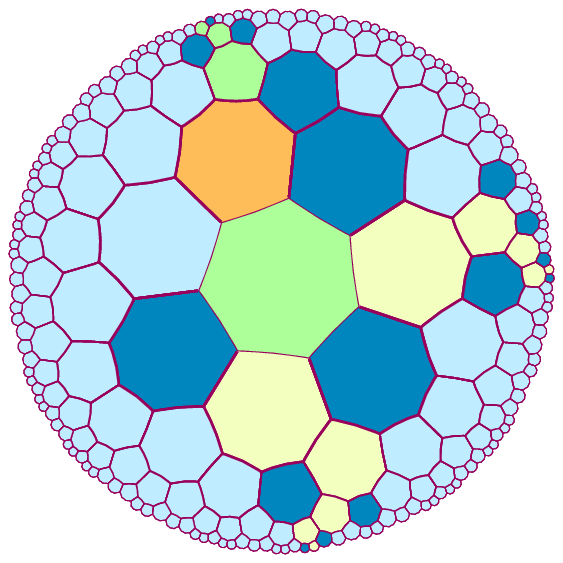}
\hfill}
\vspace{-20pt}
\ligne{\hfill
\vtop{\leftskip 0pt\parindent 0pt\hsize=300pt
\begin{fig}\label{stab_capt}
\leurre
Idle configuration of the sensor of the passive memory switch. In yellow, the
cell~$1(1)$, the sensor-cell.
\end{fig}
}
\hfill}
}

We can just note that the change of colour is different in the sensor: when the
sensor is white, if a locomotive passes, it must become black: the signal is the
locomotive itself, as will be seen in the Subsection~\ref{srcapt}. This is why the 
cell~0(0) is green in Figure~\ref{stab_capt}: the cell 1(1) can see the locomotive
only when it is in~0(0). When the sensor is 
black, it has to be changed if a locomotive passed through the other sensor which
then changed from white to black. The locomotive which arrived at the formerly white
sensor is sent to the still black one in order to make it change to white. The 
locomotive arrives through the green path of Figure~\ref{stab_capt}. As the configuration 
is the same around the cell~1(1) of that Figure as that around~1(3) in 
Figure~\ref{stab_ctrl}, the change from black to white is performed.

\section{Rules}
\label{rules}

    The figures of Section~\ref{scenar} help us to establish the rules. Their application
is illustrated by figures of this section which
were drawn by a computer program which checked the 
coherence of the rules. The program also wrote the PostScript files of the pictures from 
the computation of the application of the rules to the configurations of the various types 
of parts of the circuit. The computer program also established the traces of execution 
which contribute to the checking of the application of the rules.

    Let us explain the format of the rules and what is allowed
by the relaxation from rotation invariance. We remind the reader that a rule has the form
\hbox{\zz{$\underline{\hbox{X}}$$_o$X$_1$..X$_7$$\underline{\hbox{X}}$$_n$}},
where \hbox{\zz X$_o$} is the state of the cell~$c$,
\hbox{\zz X$_i$} is the \textbf{current} state of the neighbour~$i$ of~$c$ 
and \hbox{\zz X$_n$} is the \textbf{new} state of~$c$ applied by the
rule. As the rules no more observe the rotation invariance, we may freely 
choose which is side~1 for each cell. We take this freedom from the format of a rule 
which only requires to know which is neighbour~1. In order to restrict the number of 
rules, it is decided that, as a general rule, for a cell which is an element of the track, 
side~1 is the side shared by the cell and its next neighbour on the track, so that
tracks are one way, as already mentioned. There can be exceptions when the cell is in 
a switch or the neighbour of the central cell in a switch.
In particular, when a cell belongs to two tracks, side~1 is arbitrarily chosen among the 
two possible cases. At some places, side~1 may be chosen in order to allow the cellular
automaton to apply the expected rule. The milestones have their side~1 shared with
an element of the track which also contributes to reduce the number of rules.

    We have to keep in mind that there are two types of rules. Those
which keep the structure invariant when it is idle, we call this
type of rules \textbf{conservative}, and those which control the motion of the locomotive.
Those latter rules, which we call \textbf{motion rules}, are the rules applied to the
cells of  the tracks as well as their milestones and, sometimes to the cells of the 
structures which may be affected by the passage of the locomotive. Next, in each 
sub-section, we give the rules for the motion of the locomotive in the tracks,
then for the fixed switch, then for the doubler and for the fork, then for the selector,
then for the controller and, eventually, for the sensor. In each sub-section, we also 
illustrate the motion of the locomotive in the structure as well as a table giving traces
of execution for the cells of the track involved in the crossing.

\def\Rr#1{{\color{red}{#1}}}
\def\Yy#1{{\color{yellow}{#1}}}
\def\Gg#1{{\color{green}{#1}}}
\def\Oo#1{{\color{orange}{#1}}}
\def\Pp#1{{\color{purple}{#1}}}
\def\Bb#1{{\color{cyan}{#1}}}
\def\aff #1 #2 #3 #4 {\ligne{\hfill\footnotesize\tt\hbox to 13pt{\hfill#1}
\hskip 5pt$\underline{\hbox{\tt#2}}$#3$\underline{\hbox{\tt#4}}$\hfill}\vskip-4pt
}
\def\raff #1 #2 #3 #4 {\ligne{\hfill\footnotesize\tt\hbox to 13pt{\hfill{\Rr{#1}}}
\hskip 5pt$\underline{\hbox{\tt#2}}$#3$\underline{\hbox{\tt#4}}$\hfill}\vskip-4pt
}
\def\laff #1 #2 #3 #4 {\hbox{{\footnotesize
$\underline{\hbox{\tt#1}}${\tt#2}$\underline{\hbox{\tt#3}}${}}}#4\hskip 4pt
}
\def\haff #1 #2 #3 #4 {\hbox{\footnotesize\tt\hbox to 13pt{\hfill#1}
\hskip 5pt$\underline{\hbox{\tt#2}}$#3$\underline{\hbox{\tt#4}}$}
}
\def\hraff #1 #2 #3 #4 {\hbox{\footnotesize\tt\hbox to 13pt{\hfill{\Rr{#1}}}
\hskip 5pt$\underline{\hbox{\tt#2}}$#3$\underline{\hbox{\tt#4}}$}
}
\newdimen\tabruleli\tabruleli=300pt
\newdimen\tabrulecol\tabrulecol=60pt

\subsection{The rules for the tracks}
\label{srtrack}

    Figure~\ref{elemtrack} shows us a single element of the track. 
Figure~\ref{stab_tracks}
shows us how to assemble elements as illustrated in Figure~\ref{elemtrack} in order
to constitute tracks. In Figure~\ref{stab_tracks}, see page~\pageref{stab_tracks}.
In that figure, the tracks is represented by the yellow cells, in this order when going
from top to bottom: the cells 16(1), 6(1), 7(1), 3(1), 1(1), 
1(7), 1(6), 1(5), 1(4), 3(4), 10(4), 11(4) and 29(4).

   A close look at the tracks shows us at least two kinds of cells despite all of them
are three-milestoned cells, another difference with~\cite{mmarXiv1606} where 
four-milestoned elements of the tracks are often present. In 
Figure~\ref{stab_tracks}, there are three-milestoned cells with milestones in their
neighbours~2, 4 and~7 as, for instance, 1(6) and 3(1), three-milestoned cells with
their milestones in neighbours~3, 5 and~7, as 1(1) and~7(1) for instance. 
Table~\ref{evms} shows us that for the cells 1(6) and~3(1), rules~4, \Rr{36}, \Rr{17}, 
25 are applied, while rules~3, \Rr{38}, \Rr{40}, 43 are applied for the cells~1(1) 
and~7(1). Here and later, we write in red the number of a rule whose new state is
different from the current state of the rule.
Another assortment of the milestones is in neighbours~2, 5 and~7. Table~\ref{rmotion}
gives the motion rules corresponding to these cells and from a few others as
indicated by Tables~\ref{evms} and~\ref{evds}. The rules are borrowed from
Table~\ref{rvoies} which displays all rules used by the locomotives on the tracks.

Note that Table~\ref{rmotion} gives all possible neighbourhoods for three isolated
milestones, requiring that neighbour~1 be blank in the idle configuration. For instance,
take the positions \hbox{\tt 2, 4, 6} for the milestones. Rule~14 is the conservative
rule and rule~\Rr{31} corresponds to the case of a simple locomotive being in the cell.
Now, as the locomotive always leaves the cell through neighbour~1, rule~34 applies after
rule~\Rr{31}. Now, there are \textit{a priori} three possible entrances for the 
locomotive~:
neighbour~5 as in rule~\Rr{28} displayed in Table~\ref{rmotion}. Table~\ref{rvoies}
also shows that neighbour~3 and~7 are also used, look at rules~\Rr{63},
\laff {W} {WBBBWBW} {B} {} and~\Rr{41}, \laff {W} {WBWBWBB} {B} {.} The same observation
can be made for the other dispatches of the milestones in Table~\ref{rmotion}. In this
table, we also mark the position of the locomotive as \zz{\Bb B}.

\vskip 5pt
\ligne{\hfill
\vtop{\hsize=300pt
\begin{tab}\label{rmotion}\leurre
The motion rules for a simple locomotive.
\end{tab}
\vskip -2pt
\trep
\vskip 8pt
\ligne{\hfill
\vtop{\leftskip 0pt\parindent 0pt\hsize=\tabrulecol  
\ligne{\hskip 20pt\tt 2, 4, 6\hfill}
\aff  { 14} {W} {WBWBWBW} {W}
\raff { 28} {W} {WBWB\Bb BBW} {B}
\raff { 31} {\Bb B} {WBWBWBW} {W}
\aff  { 34} {W} {\Bb BBWBWBW} {W}
}\hskip 10pt
\vtop{\leftskip 0pt\parindent 0pt\hsize=\tabrulecol  
\ligne{\hskip 20pt\tt 2, 4, 7\hfill}
\aff  {  4} {W} {WBWBWWB} {W} 
\raff { 36} {W} {WB\Bb BBWWB} {B} 
\raff { 17} {\Bb B} {WBWBWWB} {W} 
\aff  { 25} {W} {\Bb BBWBWWB} {W}
}\hskip 10pt
\vtop{\leftskip 0pt\parindent 0pt\hsize=\tabrulecol  
\ligne{\hskip 20pt\tt 2, 5, 7\hfill}
\aff  {  7} {W} {WBWWBWB} {W}
\raff { 16} {W} {WBWWB\Bb BB} {B}
\raff { 24} {\Bb B} {WBWWBWB} {W}
\aff  { 29} {W} {\Bb BBWWBWB} {W}
}\hskip 10pt
\vtop{\leftskip 0pt\parindent 0pt\hsize=\tabrulecol  
\ligne{\hskip 20pt\tt 3, 5, 7\hfill}
\aff  {  3} {W} {WWBWBWB} {W}
\raff { 38} {W} {WWBWB\Bb BB} {B}
\raff { 40} {\Bb B} {WWBWBWB} {W}
\aff  { 43} {W} {\Bb BWBWBWB} {W}
}
\hfill}
\vskip 9pt
\trfn
\vskip 8pt
}
\hfill}
\vskip 5pt
\ligne{\hfill 
\vtop{\leftskip 0pt\parindent 0pt\hsize=\tabruleli  
\begin{tab}\label{rvoies}
\leurre
Rules managing the motion of a locomotive on the tracks.
\end{tab}
\vskip-2pt
\trep
\vskip 8pt
\ligne{\hfill from down to top : simple locomotive\hfill}
\ligne{\hfill   
\vtop{\leftskip 0pt\parindent 0pt\hsize=\tabrulecol  
\aff  {  1} {W} {WWWWWWW} {W}
\aff  {  2} {B} {WWWWWWW} {B}
\aff  {  3} {W} {WWBWBWB} {W}
\aff  {  4} {W} {WBWBWWB} {W}
\aff  {  5} {W} {WWBWWWB} {W}
\aff  {  6} {W} {BWWWWBW} {W}
\aff  {  7} {W} {WBWWBWB} {W}
\aff  {  8} {W} {WBWWWWB} {W}
\aff  {  9} {W} {WBWWWWW} {W}
\aff  { 10} {W} {WWWWWWB} {W}
\aff  { 11} {W} {WWWBWBW} {W}
\aff  { 12} {W} {BWBWWWW} {W}
}\hskip 10pt
\vtop{\leftskip 0pt\parindent 0pt\hsize=\tabrulecol  
\aff  { 13} {W} {BWWWWWW} {W}
\aff  { 14} {W} {WBWBWBW} {W}
\aff  { 15} {B} {WWBWWWW} {B}
\raff { 16} {W} {WBWWBBB} {B}
\raff { 17} {B} {WBWBWWB} {W}
\aff  { 18} {W} {BBWWWWW} {W}
\aff  { 19} {B} {BWWWWWW} {B}
\aff  { 20} {W} {BBWWWWB} {W}
\aff  { 21} {W} {WBBWWWW} {W}
\aff  { 22} {B} {WBWWWWW} {B}
\aff  { 23} {B} {WWWWWWB} {B}
\raff { 24} {B} {WBWWBWB} {W}
}\hskip 10pt
\vtop{\leftskip 0pt\parindent 0pt\hsize=\tabrulecol  
\aff  { 25} {W} {BBWBWWB} {W}
\aff  { 26} {W} {BWWWWWB} {W}
\aff  { 27} {W} {WWBWWWW} {W}
\raff { 28} {W} {WBWBBBW} {B}
\aff  { 29} {W} {BBWWBWB} {W}
\aff  { 30} {W} {BWWWWBB} {W}
\raff { 31} {B} {WBWBWBW} {W}
\raff { 32} {W} {WBBWBWB} {B}
\aff  { 33} {W} {WBBWWWB} {W}
\aff  { 34} {W} {BBWBWBW} {W}
\aff  { 35} {W} {BWBWWWB} {W}
\raff { 36} {W} {WBBBWWB} {B}
}\hskip 10pt
\vtop{\leftskip 0pt\parindent 0pt\hsize=\tabrulecol  
\aff  { 37} {B} {WWWBWWW} {B}
\raff { 38} {W} {WWBWBBB} {B}
\aff  { 39} {B} {WWWWBWW} {B}
\raff { 40} {B} {WWBWBWB} {W}
\raff { 41} {W} {WBWBWBB} {B}
\aff  { 42} {W} {BBBWWWW} {W}
\aff  { 43} {W} {BWBWBWB} {W}
\aff  { 44} {B} {WWWWWBW} {B}
\raff { 45} {W} {WBWBBWB} {B}
}
\hfill}
\vskip 8pt
\ligne{\hfill from down to top: double locomotive\hfill}
\ligne{\hfill   
\vtop{\leftskip 0pt\parindent 0pt\hsize=\tabrulecol  
\aff  { 46} {B} {WBBWWWW} {B}
\aff  { 47} {B} {WBWWBBB} {B}
\raff { 48} {B} {BBWBWWB} {W}
\aff  { 49} {B} {BBWWWWW} {B}
\aff  { 50} {B} {BWWWWWB} {B}
}\hskip 10pt
\vtop{\leftskip 0pt\parindent 0pt\hsize=\tabrulecol  
\raff { 51} {B} {BBWWBWB} {W}
\aff  { 52} {B} {WBWBBBW} {B}
\raff { 53} {B} {BBWBWBW} {W}
\aff  { 54} {B} {WBBWBWB} {B}
\aff  { 55} {W} {BBBWWWB} {W}
}\hskip 10pt
\vtop{\leftskip 0pt\parindent 0pt\hsize=\tabrulecol  
\aff  { 56} {B} {WBBBWWB} {B}
\aff  { 57} {B} {WWBBWWW} {B}
\aff  { 58} {B} {WWWBBWW} {B}
\aff  { 59} {B} {WWBWBBB} {B}
}\hskip 10pt
\vtop{\leftskip 0pt\parindent 0pt\hsize=\tabrulecol  
\raff { 60} {B} {BWBWBWB} {W}
\aff  { 61} {B} {WBWBWBB} {B}
\aff  { 62} {B} {WWWWWBB} {B}
}
\hfill}  
\vskip 8pt
\ligne{\hfill from top to bottom, when the locomotive is:\hfill}
\ligne{\hskip 40pt\hbox to \tabrulecol{\hfill simple\hfill}
\hskip 120pt double\hfill}
\ligne{\hfill
\vtop{\leftskip 0pt\parindent 0pt\hsize=\tabrulecol  
\raff { 63} {W} {WBBBWBW} {B}
}\hskip 10pt
\vtop{\leftskip 0pt\parindent 0pt\hsize=\tabrulecol  
\aff  { 64} {B} {WWWWBBW} {B}
}\hskip 10pt
\vtop{\leftskip 0pt\parindent 0pt\hsize=\tabrulecol  
\aff  { 65} {B} {WBBBWBW} {B}
}\hskip 10pt
\vtop{\leftskip 0pt\parindent 0pt\hsize=\tabrulecol  
\aff  { 66} {B} {WBWBBWB} {B}
}
\hfill}
\vskip 9pt
\trfn
\vskip 8pt
}
\hfill}

\newdimen\tabvois\tabvois=37pt
\newdimen\tabcol\tabcol=80pt

Accordingly, for the display \hbox to \tabvois {\tt 2, 4, 7}, besides rule~\Rr{36} 
corresponding to an entrance through neighbour~3, rule~\Rr{41} again corresponds to 
another one through neighbour~6, and rule~\Rr{45}
\laff {W} {WBWBBWB} {B}  {} corresponds to again another one through neighbour~5. 
This can be repeated for the other neighbours, see Table~\ref{mvflex} where the
number in brackets indicates the neighbour through which the locomotive enters.
In that table the front of the locomotive is marked in the rules as \zz{\Bb B}. 
Rule~\Rr{71} will be used later, in the fixed switch.
\vskip 5pt
\ligne{\hfill
\vtop{\leftskip 0pt\parindent 0pt\hsize= 320pt
\begin{tab}\label{mvflex}\leurre
The other rules involved for the motion of a simple locomotive.
\end{tab}
\vskip -2pt
\trep
\vskip 8pt
\ligne{\hfill
\hbox to \tabvois {\tt 3, 5, 7} : 
\hbox to \tabcol{\hraff { 38} {W} {WWBWB\Bb BB} {B} [6]}\hskip 10pt
\hbox to \tabcol{\hraff { 71} {W} {WWB\Bb BBWB} {B} [4]}\hskip 10pt 
\hbox to \tabcol{\hraff { 32} {W} {W\Bb BBWBWB} {B} [2]}\hfill}
\ligne{\hfill
\hbox to \tabvois {\tt 2, 5, 7} :
\hbox to \tabcol{\hraff { 16} {W} {WBWWB\Bb BB} {B} [6]}\hskip 10pt
\hbox to \tabcol{\hraff { 45} {W} {WBW\Bb BBWB} {B} [4]}\hskip 10pt
\hbox to \tabcol{\hraff { 32} {W} {WB\Bb BWBWB} {B} [3]}\hfill}
\ligne{\hfill
\hbox to \tabvois {\tt 2, 4, 7} : 
\hbox to \tabcol{\hraff { 36} {W} {WB\Bb BBWWB} {B} [3]}\hskip 10pt 
\hbox to \tabcol{\hraff { 41} {W} {WBWBW\Bb BB} {B} [6]}\hskip 10pt
\hbox to \tabcol{\hraff { 45} {W} {WBWB\Bb BWB} {B} [5]}\hfill}
\ligne{\hfill
\hbox to \tabvois {\tt 2, 4, 6} : 
\hbox to \tabcol{\hraff { 28} {W} {WBWB\Bb BBW} {B} [5]}\hskip 10pt
\hbox to \tabcol{\hraff { 63} {W} {WB\Bb BBWBW} {B} [3]}\hskip 10pt
\hbox to \tabcol{\hraff { 41} {W} {WBWBW\Bb BB} {B} [7]}\hfill}
\vskip 9pt
\trfn
\vskip 8pt
}
\hfill}

Call the rules of Table~\ref{rmotion} for
a given neighbourhood, the conservative rule, the front rule, the cell rule and the 
witness rule, the names being self explanatory.

Table~\ref{rmotion} also shows us an interesting features: the neighbourhood of
rule~\Rr{36} is \hbox{\zz{WBBBWWB}} and that of rule~29 is \hbox{\zz{BBWWBWB}}. It is not
difficult to see that we can pass from one neighbourhood to the other by a circular
permutations. Accordingly, rule~\Rr{36} and rule~29 are not rotationally compatible:
rule~\Rr{36} is a front rule, rule~29 is a witness one.
The other conclusion we can draw from the comments regarding other variants of
the rule allowing to make the locomotive enter the cell is that this number of
variants gives us an important flexibility for devising tracks which go from a tile to
another one. On the example of Figure~\ref{stab_tracks} we can see that the track
going to 1(6) can also go to~2(6), 3(6) or 4(6). Figure~\ref{cont_tracks} show us
how to proceed to continue a path to the sons of an already reached element.
\vskip 10pt
\vtop{
\ligne{\hfill
\includegraphics[scale=0.65]{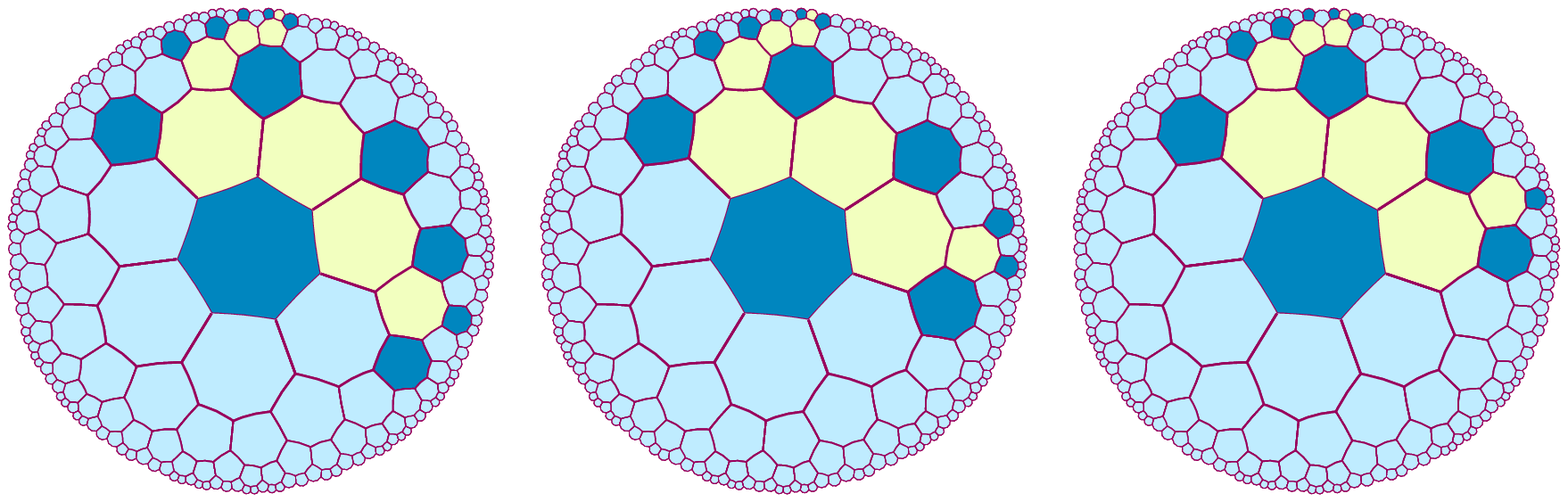}
\hfill}
\vspace{-10pt}
\ligne{\hfill
\vtop{\leftskip 0pt\parindent 0pt\hsize=200pt
\begin{fig}\label{cont_tracks}
\leurre
Element of the tracks: in yellow, the elements of the track where the locomotive
passes.
\end{fig}
}
\hfill}
}

In Figure~\ref{stab_tracks}, the neighbourhood of 1(6) is of the type
\hbox to \tabvois{\tt 2, 5, 7}. On the pictures of Figure~\ref{cont_tracks}, we have the following
neighbourhoods:
\vskip 5pt
\ligne{\hfill
\vtop{\leftskip 0pt\parindent 0pt\hsize300pt
\ligne{\hfill \hbox to 20pt{1(6):}\hfill
\hbox to \tabvois {\tt 2, 4, 6}
\hskip 20pt
\hbox to \tabvois {\tt 3, 5, 7}
\hskip 20pt
\hbox to \tabvois {\tt 2, 4, 7}\hfill}
\ligne{\hfill \hbox to 20pt{sons):}\hfill
\hbox to \tabvois {\hfill 2(6)\hfill}
\hskip 20pt
\hbox to \tabvois {\hfill 3(6)\hfill}
\hskip 20pt
\hbox to \tabvois {\hfill 4(6)\hfill}\hfill}
\ligne{\hfill \hbox to 20pt{\hfill}\hfill
\hbox to \tabvois {\tt 2, 4, 6}
\hskip 20pt
\hbox to \tabvois {\tt 2, 5, 7}
\hskip 20pt
\hbox to \tabvois {\tt 2, 5, 7}\hfill}
}
\hfill}
\vskip 5pt

It is worth noticing that many rules appearing in Table~\ref{evms} showing the rules
used when a simple locomotive goes up along the tracks illustrated by 
Figure~\ref{stab_tracks} also appear in Table~\ref{evds} which display the rules used
when the same locomotive goes down, assuming that the sides~1 have been changed in order
to allow the motion from top to bottom on the same cells. Of course, in the circuit,
for any cell, side~1 is fixed once and for all.

\newdimen\largez\largez=20pt
\def\HH#1{\hbox to\largez{\hfill #1\hfill}}
\ligne{\hfill
\vtop{\hsize=260pt
\begin{tab}\label{evms}\leurre
Execution of the rules $1$ up to~$45$: motion along the tracks from down to top
for a simple locomotive.
\end{tab}
\vskip 2pt
\trep
\vskip 8pt
\ligne{\hfill\HH{}
\HH{{11$_4$}}\HH{{10$_4$}}\HH{{3$_4$} }\HH{{1$_4$} }\HH{{1 $_5$}}\HH{{1$_6$} }\HH{{1$_7$} }\HH{{1$_1$} }\HH{{3$_1$} }\HH{{7$_1$} }\HH{{6$_1$} }
\hfill}
\ligne{\hfill\HH{1}
\HH{25}\HH{\Rr{24}}\HH{\Rr{16}}\HH{14}\HH{7}\HH{4}\HH{4}\HH{3}\HH{4}\HH{3}\HH{7}\hfill}
\ligne{\hfill\HH{2}
\HH{4}\HH{29}\HH{\Rr{24}}\HH{\Rr{28}}\HH{7}\HH{4}\HH{4}\HH{3}\HH{4}\HH{3}\HH{7}\hfill}
\ligne{\hfill\HH{3}
\HH{4}\HH{7}\HH{29}\HH{\Rr{31}}\HH{\Rr{32}}\HH{4}\HH{4}\HH{3}\HH{4}\HH{3}\HH{7}\hfill}
\ligne{\hfill\HH{4}
\HH{4}\HH{7}\HH{7}\HH{34}\HH{\Rr{24}}\HH{\Rr{36}}\HH{4}\HH{3}\HH{4}\HH{3}\HH{7}\hfill}
\ligne{\hfill\HH{5}
\HH{4}\HH{7}\HH{7}\HH{14}\HH{29}\HH{\Rr{17}}\HH{\Rr{36}}\HH{3}\HH{4}\HH{3}\HH{7}\hfill}
\ligne{\hfill\HH{6}
\HH{4}\HH{7}\HH{7}\HH{14}\HH{7}\HH{25}\HH{\Rr{17}}\HH{\Rr{38}}\HH{4}\HH{3}\HH{7}\hfill}
\ligne{\hfill\HH{7}
\HH{4}\HH{7}\HH{7}\HH{14}\HH{7}\HH{4}\HH{25}\HH{\Rr{40}}\HH{\Rr{41}}\HH{3}\HH{7}\hfill}
\ligne{\hfill\HH{8}
\HH{4}\HH{7}\HH{7}\HH{14}\HH{7}\HH{4}\HH{4}\HH{43}\HH{\Rr{17}}\HH{\Rr{38}}\HH{7}\hfill}
\ligne{\hfill\HH{9}
\HH{4}\HH{7}\HH{7}\HH{14}\HH{7}\HH{4}\HH{4}\HH{3}\HH{25}\HH{\Rr{40}}\HH{\Rr{45}}\hfill}
\vskip 9pt
\trfn
\vskip 15pt
}
\hfill}

\ligne{\hfill
\vtop{\hsize=260pt
\begin{tab}\label{evds}\leurre
Execution of the rules $1$ up to~$45$: motion along the tracks from top to bottom
for a simple locomotive.
\end{tab}
\vskip 2pt
\trep
\vskip 8pt
\ligne{\hfill\HH{}
\HH{{6$_1$} }\HH{{7$_1$} }\HH{{3$_1$} }\HH{{1$_1$} }\HH{{1$_7$} }\HH{{1$_6$} }\HH{{1$_5$} }\HH{{1$_4$} }\HH{{3$_4$} }\HH{{10$_4$}}\HH{{11$_4$}}
\hfill}
\ligne{\hfill\HH{1}
\HH{34}\HH{\Rr{24}}\HH{\Rr{63}}\HH{7}\HH{7}\HH{7}\HH{3}\HH{7}\HH{4}\HH{4}\HH{7}\hfill}
\ligne{\hfill\HH{2}
\HH{14}\HH{29}\HH{\Rr{31}}\HH{\Rr{32}}\HH{7}\HH{7}\HH{3}\HH{7}\HH{4}\HH{4}\HH{7}\hfill}
\ligne{\hfill\HH{3}
\HH{14}\HH{7}\HH{34}\HH{\Rr{24}}\HH{\Rr{16}}\HH{7}\HH{3}\HH{7}\HH{4}\HH{4}\HH{7}\hfill}
\ligne{\hfill\HH{4}
\HH{14}\HH{7}\HH{14}\HH{29}\HH{\Rr{24}}\HH{\Rr{16}}\HH{3}\HH{7}\HH{4}\HH{4}\HH{7}\hfill}
\ligne{\hfill\HH{5}
\HH{14}\HH{7}\HH{14}\HH{7}\HH{29}\HH{\Rr{24}}\HH{\Rr{38}}\HH{7}\HH{4}\HH{4}\HH{7}\hfill}
\ligne{\hfill\HH{6}
\HH{14}\HH{7}\HH{14}\HH{7}\HH{7}\HH{29}\HH{\Rr{40}}\HH{\Rr{45}}\HH{4}\HH{4}\HH{7}\hfill}
\ligne{\hfill\HH{7}
\HH{14}\HH{7}\HH{14}\HH{7}\HH{7}\HH{7}\HH{43}\HH{\Rr{24}}\HH{\Rr{36}}\HH{4}\HH{7}\hfill}
\ligne{\hfill\HH{8}
\HH{14}\HH{7}\HH{14}\HH{7}\HH{7}\HH{7}\HH{3}\HH{29}\HH{\Rr{17}}\HH{\Rr{36}}\HH{7}\hfill}
\vskip 9pt
\trfn
\vskip 15pt
}
\hfill}
 Before turning to the rules for a double locomotive, note that in the conservative rules
for the elements of the tracks the neighbourhoods are rotated forms of each other.

The rules for the double locomotive are displayed in Table~\ref{rvoies}. They can be 
derived from the previous rules as follows. The conservative and the front rules are 
the same: at those times, the cell do not know whether the locomotive is simple or double.
Once the front is in the cell, the cell rule cannot be applied as the rear is seen at the
place were the front was by one time backwards. Accordingly the cell rule is replaced
by two new rules: the rear rule, which makes the second cell of the locomotive enter
the cell and the clearing rule which makes it leave the cell. Note that the
rear, clearing rules are obtained from the front, witness rules respectively by 
changing the current state from \zz W to~\zz B. Table~\ref{rmotiond} gives the rules
applied in the respective neighbourhoods considered in Table~\ref{rmotion}.
That table has the same property as Table~\ref{rmotion}. Other entries are possible for 
the double locomotive.  Table~\ref{mvflex} indicates us for each neighbourhood the
sequence of rules constituted, in this order, by the front, the rear and the clearing 
rules. Also, brackets indicate, after each triple, which is the neighbour through which 
the locomotive enters the cell. In that table, the locomotive is marked as
\zz{\Bb B}. In Table~\ref{rmotiond}, the front of the locomotive is \zz{\Bb B} and its
rear is \zz{\Oo B}. Table~\ref{mvdflex} indicates all possible neighbourhoodes with,
in brackets the entry of the front of the locomotive. 
Note the same phenomenon as in Table~\ref{rmotion}: 
a few pair of rear and clearing rules are the same for different neighbours. In fact that
feature appears when two neighbours differ by one place, for example
\hbox to \tabvois {\tt 2, 4, 7} and \hbox to \tabvois {\tt 2, 5, 7}. Clearly,
the positions of the black neighbours is the same for the front and rear rules
when the entrance is neighbour~5 and~4 respectively.

\vskip 5pt
\ligne{\hfill
\vtop{\hsize=320pt
\begin{tab}\label{rmotiond}\leurre
The motion rules for a double locomotive.
\end{tab}
\vskip -2pt
\trep
\vskip 8pt
\ligne{\hfill
\vtop{\leftskip 0pt\parindent 0pt\hsize=\tabrulecol  
\ligne{\hskip 20pt\tt 2, 4, 6\hfill}
\aff  { 14} {W} {WBWBWBW} {W}
\raff { 28} {W} {WBWB\Bb BBW} {B}  
\aff  { 52} {\Bb B} {WBWB\Oo BBW} {B}
\raff { 53} {\Oo B} {\Bb BBWBWBW} {W}
\aff  { 34} {W} {\Oo BBWBWBW} {W}
}\hskip 10pt
\vtop{\leftskip 0pt\parindent 0pt\hsize=\tabrulecol  
\ligne{\hskip 20pt\tt 2, 4, 7\hfill}
\aff  {  4} {W} {WBWBWWB} {W} 
\raff { 36} {W} {WB\Bb BBWWB} {B}  
\aff  { 56} {\Bb B} {WB\Oo BBWWB} {B}
\raff { 48} {\Oo B} {\Bb BBWBWWB} {W}
\aff  { 25} {W} {\Oo BBWBWWB} {W}
}\hskip 10pt
\vtop{\leftskip 0pt\parindent 0pt\hsize=\tabrulecol  
\ligne{\hskip 20pt\tt 2, 5, 7\hfill}
\aff  {  7} {W} {WBWWBWB} {W}
\raff { 16} {W} {WBWWB\Bb BB} {B}
\aff  { 47} {\Bb B} {WBWWB\Oo BB} {B}
\raff { 51} {\Oo B} {\Bb BBWWBWB} {W}
\aff  { 29} {W} {\Oo BBWWBWB} {W}
}\hskip 10pt
\vtop{\leftskip 0pt\parindent 0pt\hsize=\tabrulecol  
\ligne{\hskip 20pt\tt 3, 5, 7\hfill}
\aff  {  3} {W} {WWBWBWB} {W}
\raff { 38} {W} {WWBWB\Bb BB} {B}
\aff  { 59} {\Bb B} {WWBWB\Oo BB} {B}
\raff { 60} {\Oo B} {BWBWBWB} {W}
\aff  { 43} {W} {\Oo BWBWBWB} {W}
}
\hfill}
\vskip 9pt
\trfn
\vskip 8pt
}
\hfill}
\vskip 5pt
\ligne{\hfill
\vtop{\leftskip 0pt\parindent 0pt\hsize= 300pt
\begin{tab}\label{mvdflex}\leurre
The other rules involved for the motion of a double locomotive.
\end{tab}
\vskip -2pt
\trep
\vskip 8pt
\ligne{\hfill
\hbox to \tabvois {\tt 3, 5, 7} : 
\hbox to \tabrulecol{\Rr{38}, 59, \Rr{60} [6]}\hskip 10pt
\hbox to \tabrulecol{\Rr{71}, 72, \Rr{60} [4]}\hskip 10pt 
\hbox to \tabrulecol{\Rr{32}, 54, \Rr{60} [2]}\hfill}
\ligne{\hfill
\hbox to \tabvois {\tt 2, 5, 7} :
\hbox to \tabrulecol{\Rr{16}, 47, \Rr{51} [6]}\hskip 10pt
\hbox to \tabrulecol{\Rr{45}, 66, \Rr{51} [4]}\hskip 10pt
\hbox to \tabrulecol{\Rr{32}, 53, \Rr{51} [3]}\hfill}
\ligne{\hfill
\hbox to \tabvois {\tt 2, 4, 7} : 
\hbox to \tabrulecol{\Rr{36}, 56, \Rr{48} [3]}\hskip 10pt 
\hbox to \tabrulecol{\Rr{41}, 61, \Rr{48} [6]}\hskip 10pt
\hbox to \tabrulecol{\Rr{45}, 67, \Rr{48} [5]}\hfill}
\ligne{\hfill
\hbox to \tabvois {\tt 2, 4, 6} : 
\hbox to \tabrulecol{\Rr{28}, 52, \Rr{53} [5]}\hskip 10pt
\hbox to \tabrulecol{\Rr{63}, 65, \Rr{53} [3]}\hskip 10pt
\hbox to \tabrulecol{\Rr{41}, 61, \Rr{53} [7]}\hfill}
\vskip 9pt
\trfn
\vskip 8pt
}
\hfill}

Table~\ref{rmotiond} contains 20 rules. The cell rule is different in Table~\ref{rmotion}
so that, together, those tables contain 24 rules. Table~\ref{mvdflex} brings in
16 new rules. Accordingly we have 40 rules for the elements of the track only.
There are other rules concerning the milestones: \laff {B} {W$^\alpha$BW$^\beta$} {B} {}
with \hbox{$\alpha$,$\beta \geq 0$} and \hbox{$\alpha$+$\beta = 6$} together
with \hbox{$\alpha<5$} when a simple locomotive moves, see rules~19, 22, 15, 37 and 39.
For a double one, we have all possible rules of the form
\laff {B} {W$^\alpha$BBW$^\beta$} {B} {}
with \hbox{$\alpha$,$\beta \geq 0$} and \hbox{$\alpha$+$\beta = 5$} together
with \laff {B} {BW$^5$B} {B} {,} see rules 49, 46, 57, 58, 64, 62 and 50. We have also
to append rule~1, \laff {W} {W$^7$} {W} {,} the conservative rule of the blank cells
which have no black cell among their neighbours, as well as rule~2, 
\laff {B} {W$^7$} {B} {,} which is the conservative rule of the milestones of the elements
of the track. Table~\ref{rvwit} illustrates the use of many of those rules together
with some others for four cells: 4(1), a white cell, and three milestones: 2(1), 0(0)
and~4(4). The cell 4(1) illustrates a situation when a white cell can see two consecutive
elements of the tracks. We use the same colour conventions as in Tables~\ref{rmotion}
and~\ref{rmotiond} for the front and for the rear of a locomotive.

It can be noted that the rules of Table~\ref{rvwit} do not change the current state
of the cell: it is conformal with the role of witness devoted to these cells. We also can 
remark that the change of direction in the motion boils down to change the order of
application of the rules. We also can see the change in the display of the colours in
the rules attached to the motion of a double locomotive.

\ligne{\hfill
\vtop{\leftskip 0pt\parindent 0pt\hsize=340pt
\begin{tab}\label{rvwit}\leurre
Rules for cells witnessing the motion on the tracks.
\end{tab}
\vskip-2pt 
\trep
\vskip 8pt
\ligne{4(1):\hfill
\vtop{\leftskip 0pt\parindent 0pt\hsize=\tabrulecol
\ligne{\hfill simple $\uparrow$\hfill}
\aff  {  5} {W} {WWBWWWB} {W}
\aff  { 35} {W} {\Bb BWBWWWB} {W}
\aff  { 33} {W} {W\Bb BBWWWB} {W}
\aff  {  5} {W} {WWBWWWB} {W}
}\hskip 10pt
\vtop{\leftskip 0pt\parindent 0pt\hsize=\tabrulecol
\ligne{\hfill double $\uparrow$\hfill}
\aff  {  5} {W} {WWBWWWB} {W}
\aff  { 35} {W} {\Bb BWBWWWB} {W}
\aff  { 55} {W} {\Oo B\Bb BBWWWB} {W}
\aff  { 33} {W} {W\Oo BBWWWB} {W}
\aff  {  5} {W} {WWBWWWB} {W}
}\hskip 10pt
\vtop{\leftskip 0pt\parindent 0pt\hsize=\tabrulecol
\ligne{\hfill simple $\downarrow$\hfill}
\aff  {  5} {W} {WWBWWWB} {W}
\aff  { 33} {W} {W\Bb BBWWWB} {W}
\aff  { 35} {W} {\Bb BWBWWWB} {W}
\aff  {  5} {W} {WWBWWWB} {W}
}\hskip 10pt
\vtop{\leftskip 0pt\parindent 0pt\hsize=\tabrulecol
\ligne{\hfill double $\downarrow$\hfill}
\aff  {  5} {W} {WWBWWWB} {W}
\aff  { 33} {W} {W\Bb BBWWWB} {W}
\aff  { 55} {W} {\Bb B\Oo BBWWWB} {W}
\aff  { 35} {W} {\Oo BWBWWWB} {W}
\aff  {  5} {W} {WWBWWWB} {W}
}
\hfill}
\vskip 5pt
\ligne{2(1):\hfill
\vtop{\leftskip 0pt\parindent 0pt\hsize=\tabrulecol
\ligne{\hfill simple $\uparrow$\hfill}  
\aff  {  2} {B} {WWWWWWW} {B}
\aff  { 22} {B} {W\Bb BWWWWW} {B}
\aff  { 19} {B} {\Bb BWWWWWW} {B}
\aff  { 23} {B} {WWWWWW\Bb B} {B}
\aff  { 44} {B} {WWWWW\Bb BW} {B}
\aff  { 39} {B} {WWWW\Bb BWW} {B}
\aff  {  2} {B} {WWWWWWW} {B}
}\hskip 10pt
\vtop{\leftskip 0pt\parindent 0pt\hsize=\tabrulecol
\ligne{\hfill double $\uparrow$\hfill}
\aff  {  2} {B} {WWWWWWW} {B}
\aff  { 22} {B} {W\Bb BWWWWW} {B}
\aff  { 49} {B} {\Bb B\Oo BWWWWW} {B}
\aff  { 50} {B} {\Oo BWWWWW\Bb B} {B}
\aff  { 62} {B} {WWWWW\Bb B\Oo B} {B}
\aff  { 64} {B} {WWWW\Bb B\Oo BW} {B}
\aff  { 39} {B} {WWWW\Oo BWW} {B}
\aff  {  2} {B} {WWWWWWW} {B}
}\hskip 10pt
\vtop{\leftskip 0pt\parindent 0pt\hsize=\tabrulecol
\ligne{\hfill simple $\downarrow$\hfill}
\aff  {  2} {B} {WWWWWWW} {B}
\aff  { 39} {B} {WWWW\Bb BWW} {B}
\aff  { 44} {B} {WWWWW\Bb BW} {B}
\aff  { 23} {B} {WWWWWW\Bb B} {B}
\aff  { 19} {B} {\Bb BWWWWWW} {B}
\aff  { 22} {B} {W\Bb BWWWWW} {B}
\aff  {  2} {B} {WWWWWWW} {B}
}\hskip 10pt
\vtop{\leftskip 0pt\parindent 0pt\hsize=\tabrulecol
\ligne{\hfill double $\downarrow$\hfill}
\aff  {  2} {B} {WWWWWWW} {B}
\aff  { 39} {B} {WWWW\Bb BWW} {B}
\aff  { 64} {B} {WWWW\Oo B\Bb BW} {B}
\aff  { 62} {B} {WWWWW\Oo B\Bb B} {B}
\aff  { 50} {B} {\Bb BWWWWW\Oo B} {B}
\aff  { 49} {B} {\Oo B\Bb BWWWWW} {B}
\aff  { 22} {B} {W\Oo BWWWWW} {B}
\aff  {  2} {B} {WWWWWWW} {B}
}
\hfill}
\vskip 5pt
\ligne{0(0):\hfill
\vtop{\leftskip 0pt\parindent 0pt\hsize=\tabrulecol
\ligne{\hfill simple $\uparrow$\hfill}  
\aff  {  2} {B} {WWWWWWW} {B}
\aff  { 19} {B} {\Bb BWWWWWW} {B}
\aff  { 22} {B} {W\Bb BWWWWW} {B}
\aff  { 15} {B} {WW\Bb BWWWW} {B}
\aff  { 37} {B} {WWW\Bb BWWW} {B}
\aff  { 39} {B} {WWWW\Bb BWW} {B}
\aff  {  2} {B} {WWWWWWW} {B}
}\hskip 10pt
\vtop{\leftskip 0pt\parindent 0pt\hsize=\tabrulecol
\ligne{\hfill double $\uparrow$\hfill}
\aff  {  2} {B} {WWWWWWW} {B}
\aff  { 19} {B} {\Bb BWWWWWW} {B}
\aff  { 49} {B} {\Oo B\Bb BWWWWW} {B}
\aff  { 46} {B} {W\Oo B\Bb BWWWW} {B}
\aff  { 57} {B} {WW\Oo B\Bb BWWW} {B}
\aff  { 58} {B} {WWW\Oo B\Bb BWW} {B}
\aff  { 39} {B} {WWWW\Oo BWW} {B}
\aff  {  2} {B} {WWWWWWW} {B}
}\hskip 10pt
\vtop{\leftskip 0pt\parindent 0pt\hsize=\tabrulecol
\ligne{\hfill simple $\downarrow$\hfill}
\aff  {  2} {B} {WWWWWWW} {B}
\aff  { 39} {B} {WWWW\Bb BWW} {B}
\aff  { 37} {B} {WWW\Bb BWWW} {B}
\aff  { 15} {B} {WW\Bb BWWWW} {B}
\aff  { 22} {B} {W\Bb BWWWWW} {B}
\aff  { 19} {B} {\Bb BWWWWWW} {B}
\aff  {  2} {B} {WWWWWWW} {B}
}\hskip 10pt
\vtop{\leftskip 0pt\parindent 0pt\hsize=\tabrulecol
\ligne{\hfill double $\downarrow$\hfill}
\aff  {  2} {B} {WWWWWWW} {B}
\aff  { 39} {B} {WWWW\Bb BWW} {B}
\aff  { 58} {B} {WWW\Bb B\Oo BWW} {B}
\aff  { 57} {B} {WW\Bb B\Oo BWWW} {B}
\aff  { 46} {B} {W\Bb B\Oo BWWWW} {B}
\aff  { 49} {B} {\Bb B\Oo BWWWWW} {B}
\aff  { 19} {B} {\Oo BWWWWWW} {B}
\aff  {  2} {B} {WWWWWWW} {B}
}
\hfill}
\vskip 5pt
\ligne{4(4):\hfill
\vtop{\leftskip 0pt\parindent 0pt\hsize=\tabrulecol
\ligne{\hfill simple $\uparrow$\hfill}  
\aff  {  2} {B} {WWWWWWW} {B}
\aff  { 15} {B} {WW\Bb BWWWW} {B}
\aff  { 22} {B} {W\Bb BWWWWW} {B}
\aff  { 19} {B} {\Bb BWWWWWW} {B}
\aff  { 23} {B} {WWWWWW\Bb B} {B}
\aff  {  2} {B} {WWWWWWW} {B}
}\hskip 10pt
\vtop{\leftskip 0pt\parindent 0pt\hsize=\tabrulecol
\ligne{\hfill double $\uparrow$\hfill}
\aff  {  2} {B} {WWWWWWW} {B}
\aff  { 15} {B} {WW\Bb BWWWW} {B}
\aff  { 46} {B} {W\Bb B\Oo BWWWW} {B}
\aff  { 49} {B} {\Bb B\Oo BWWWWW} {B}
\aff  { 50} {B} {\Oo BWWWWW\Bb B} {B}
\aff  { 23} {B} {WWWWWW\Oo B} {B}
\aff  {  2} {B} {WWWWWWW} {B}
}\hskip 10pt
\vtop{\leftskip 0pt\parindent 0pt\hsize=\tabrulecol
\ligne{\hfill simple $\downarrow$\hfill}
\aff  {  2} {B} {WWWWWWW} {B}
\aff  { 23} {B} {WWWWWW\Bb B} {B}
\aff  { 19} {B} {\Bb BWWWWWW} {B}
\aff  { 22} {B} {W\Bb BWWWWW} {B}
\aff  { 15} {B} {WW\Bb BWWWW} {B}
\aff  {  2} {B} {WWWWWWW} {B}
}\hskip 10pt
\vtop{\leftskip 0pt\parindent 0pt\hsize=\tabrulecol
\ligne{\hfill double $\downarrow$\hfill}
\aff  {  2} {B} {WWWWWWW} {B}
\aff  { 23} {B} {WWWWWW\Bb B} {B}
\aff  { 50} {B} {\Bb BWWWWW\Oo B} {B}
\aff  { 49} {B} {\Oo B\Bb BWWWWW} {B}
\aff  { 46} {B} {W\Oo B\Bb BWWWW} {B}
\aff  { 15} {B} {WW\Oo BWWWW} {B}
\aff  {  2} {B} {WWWWWWW} {B}
}
\hfill}
\vskip 9pt
\trfn
}
\hfill}
\vskip 10pt
Tables~\ref{evmd} and~\ref{evdd} show which instructions are applied to the cells
of the track when a double locomotive passes: from down to top in Table~\ref{evmd},
from top to bottom in Table~\ref{evdd}. Figure~\ref{fvdu} illustrates the motion of 
both types of locomotives when they go from down to top and Figure~\ref{fvud} does 
the same when they go from top to bottom. 

We conclude this section by a remark: in~\cite{mmarXiv1605b} and in~\cite{mmarXiv1606},
the tracks were implemented by using both three- and four-milestoned cells as elements
of the tracks. Here we succeeded to use three-milestoned cells only. The important
number of motion rules allowed us to assemble such the elements in very efficient 
structures. Also note the importance of the choice of side~1. As an example,
for the cell 4(4), its side~1 is shared with 3(4).

\ligne{\hfill
\vtop{\hsize=260pt
\begin{tab}\label{evmd}\leurre
Execution of the rules $1$ up to~$66$: the double locomotive on the tracks from
down to top.
\end{tab}
\vskip-2pt
\trep
\vskip 8pt
\ligne{\hfill\HH{}
\HH{{11$_4$}}\HH{{10$_4$}}\HH{{3$_4$} }\HH{{1$_4$} }\HH{{1 $_5$}}\HH{{1$_6$} }\HH{{1$_7$} }\HH{{1$_1$} }\HH{{3$_1$} }\HH{{7$_1$} }\HH{{6$_1$} }
\hfill}
\ligne{\hfill\HH{1}
\HH{25}\HH{\Rr{51}}\HH{47}\HH{\Rr{28}}\HH{7}\HH{4}\HH{4}\HH{3}\HH{4}\HH{3}\HH{7}\hfill}
\ligne{\hfill\HH{2}
\HH{4}\HH{29}\HH{\Rr{51}}\HH{52}\HH{\Rr{32}}\HH{4}\HH{4}\HH{3}\HH{4}\HH{3}\HH{7}\hfill}
\ligne{\hfill\HH{3}
\HH{4}\HH{7}\HH{29}\HH{\Rr{53}}\HH{54}\HH{\Rr{36}}\HH{4}\HH{3}\HH{4}\HH{3}\HH{7}\hfill}
\ligne{\hfill\HH{4}
\HH{4}\HH{7}\HH{7}\HH{34}\HH{\Rr{51}}\HH{56}\HH{\Rr{36}}\HH{3}\HH{4}\HH{3}\HH{7}\hfill}
\ligne{\hfill\HH{5}
\HH{4}\HH{7}\HH{7}\HH{14}\HH{29}\HH{\Rr{48}}\HH{56}\HH{\Rr{38}}\HH{4}\HH{3}\HH{7}\hfill}
\ligne{\hfill\HH{6}
\HH{4}\HH{7}\HH{7}\HH{14}\HH{7}\HH{25}\HH{\Rr{48}}\HH{59}\HH{\Rr{41}}\HH{3}\HH{7}\hfill}
\ligne{\hfill\HH{7}
\HH{4}\HH{7}\HH{7}\HH{14}\HH{7}\HH{4}\HH{25}\HH{\Rr{60}}\HH{61}\HH{\Rr{38}}\HH{7}\hfill}
\ligne{\hfill\HH{8}
\HH{4}\HH{7}\HH{7}\HH{14}\HH{7}\HH{4}\HH{4}\HH{43}\HH{\Rr{48}}\HH{59}\HH{\Rr{45}}\hfill}
\vskip 9pt
\trfn
\vskip 15pt
}
\hfill}

\ligne{\hfill
\vtop{\hsize=260pt
\begin{tab}\label{evdd}\leurre
Execution of the rules $1$ up to~$66$: the double locomotive on the tracks from
top to bottom.
\end{tab}
\vskip-2pt
\trep
\vskip 8pt
\ligne{\hfill\HH{}
\HH{{6$_1$} }\HH{{7$_1$} }\HH{{3$_1$} }\HH{{1$_1$} }\HH{{1$_7$} }\HH{{1$_6$} }\HH{{1$_5$} }\HH{{1$_4$} }\HH{{3$_4$} }\HH{{10$_4$}}\HH{{11$_4$}}
\hfill}
\ligne{\hfill\HH{1}
\HH{34}\HH{\Rr{51}}\HH{65}\HH{\Rr{32}}\HH{7}\HH{7}\HH{3}\HH{7}\HH{4}\HH{4}\HH{7}\hfill}
\ligne{\hfill\HH{2}
\HH{14}\HH{29}\HH{\Rr{53}}\HH{54}\HH{\Rr{16}}\HH{7}\HH{3}\HH{7}\HH{4}\HH{4}\HH{7}\hfill}
\ligne{\hfill\HH{3}
\HH{14}\HH{7}\HH{34}\HH{\Rr{51}}\HH{47}\HH{\Rr{16}}\HH{3}\HH{7}\HH{4}\HH{4}\HH{7}\hfill}
\ligne{\hfill\HH{4}
\HH{14}\HH{7}\HH{14}\HH{29}\HH{\Rr{51}}\HH{47}\HH{\Rr{38}}\HH{7}\HH{4}\HH{4}\HH{7}\hfill}
\ligne{\hfill\HH{5}
\HH{14}\HH{7}\HH{14}\HH{7}\HH{29}\HH{\Rr{51}}\HH{59}\HH{\Rr{45}}\HH{4}\HH{4}\HH{7}\hfill}
\ligne{\hfill\HH{6}
\HH{14}\HH{7}\HH{14}\HH{7}\HH{7}\HH{29}\HH{\Rr{60}}\HH{66}\HH{\Rr{36}}\HH{4}\HH{7}\hfill}
\ligne{\hfill\HH{7}
\HH{14}\HH{7}\HH{14}\HH{7}\HH{7}\HH{7}\HH{43}\HH{\Rr{51}}\HH{56}\HH{\Rr{36}}\HH{7}\hfill}
\ligne{\hfill\HH{8}
\HH{14}\HH{7}\HH{14}\HH{7}\HH{7}\HH{7}\HH{3}\HH{29}\HH{\Rr{48}}\HH{56}\HH{\Rr{16}}\hfill}
\vskip 9pt
\trfn
\vskip 15pt
}
\hfill}

As mentioned in the introduction, the pictures which constitute Figures~\ref{fvdu}
and~\ref{fvud} are produced by PostScript programs computed by a computer program.
The program applies the rules given in Table~\ref{rvoies} and the other tables 
of rules given in the following sub sections to the cellular automaton. From those
calculations, it computes the position of the cell(s) representing the locomotive on 
the tracks. The same program did the same for the various configurations we shall
further investigate.

\vtop{
\ligne{\hfill
\includegraphics[scale=0.55]{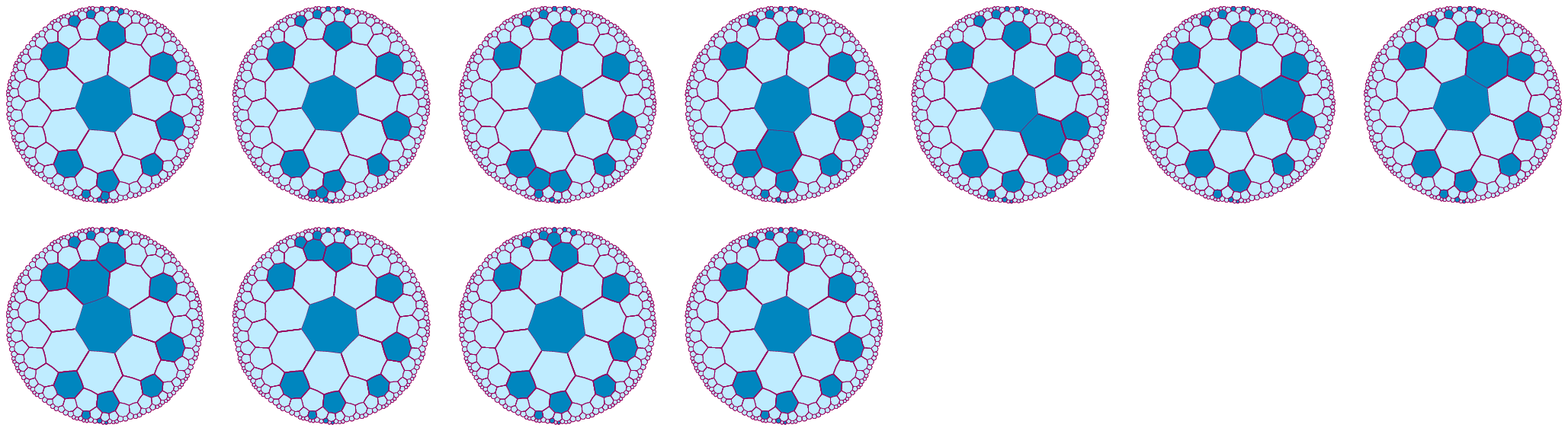}
\hfill}
\vspace{-20pt}
\ligne{\hfill
\includegraphics[scale=0.55]{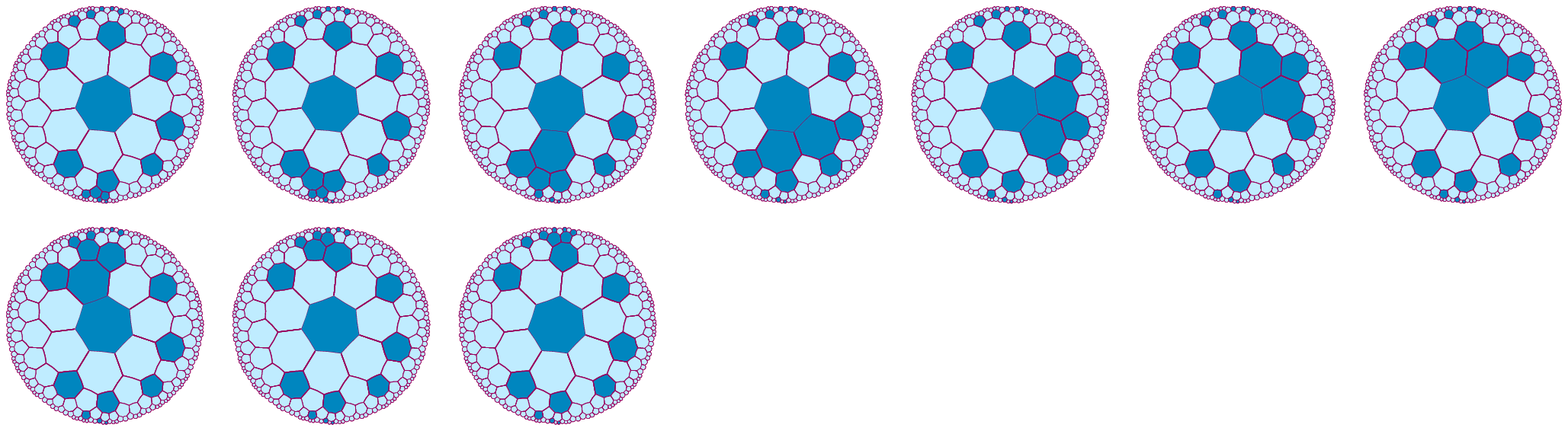}
\hfill}
\vspace{-15pt}
\ligne{\hfill
\vtop{\leftskip 0pt\parindent 0pt\hsize=300pt
\begin{fig}\label{fvdu}
\leurre
Illustration of the motion from down to top: above, for a simple locomotive,
below, for a double locomotive.
\end{fig}
}
\hfill}
}
\vskip -10pt
\vtop{
\ligne{\hfill
\includegraphics[scale=0.55]{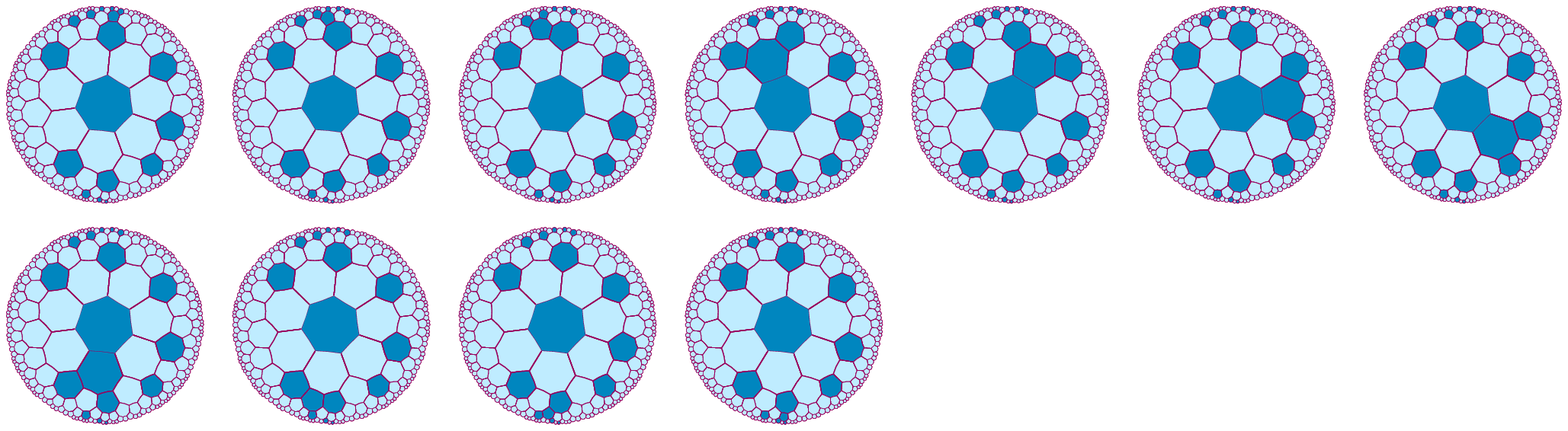}
\hfill}
\vspace{-20pt}
\ligne{\hfill
\includegraphics[scale=0.55]{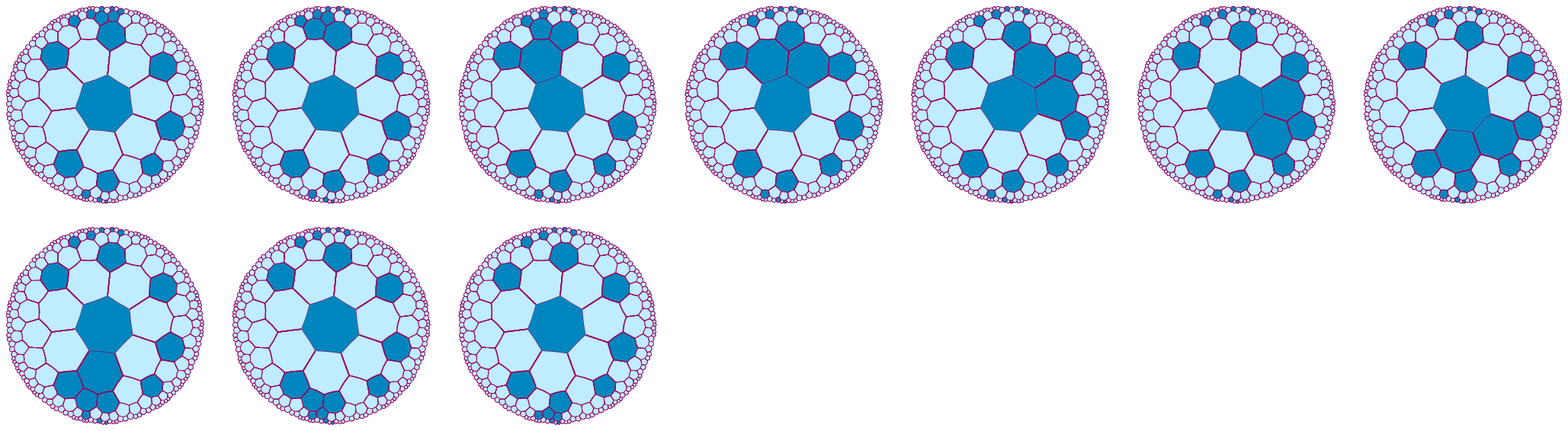}
\hfill}
\vspace{-15pt}
\ligne{\hfill
\vtop{\leftskip 0pt\parindent 0pt\hsize=300pt
\begin{fig}\label{fvud}
\leurre
Illustration of the motion from top to bottom: above, for a simple locomotive,
below, for a double locomotive.
\end{fig}
}
\hfill}
}

\subsection{The rules for the fixed switch, for the fork and for the doubler}
\label{srfxdblfrk}

We now turn to the study of the fixed switch, of the fork and of the doubler. 
Table~\ref{rfixed} gives new rules which are used for the 
crossing of those structures together with already used rules. We start our study 
with the fixed switch which is a passive structure as noted in Subsection~\ref{struct}. 

\subsubsection{The fixed switch}
\label{srfx}

As can already be seen on Figure~\ref{stab_fx}, the structure is mainly constituted
by elements of the tracks assembled in a suited way. In particular, the central
cell is a three-milestoned cell as in the elements of the tracks. Its neighbourhood is
a rotated image of any neighbourhood of a cell of the tracks as studied 
in Subsection~\ref{srtrack}.

\ligne{\hfill 
\vtop{\leftskip 0pt\parindent 0pt\hsize=\tabruleli  
\begin{tab}\label{rfixed}
\leurre
Rules for the crossing of a fixed switch, of a doubler and of a fork.
\end{tab}
\vskip-2pt
\trfn
\vskip 8pt
\ligne{fixed:\hfill \hbox to 130pt{\hfill from the left\hfill}
\hskip 10pt\hbox to 150pt{\hfill from the right\hfill}\hfill}
\ligne{\hfill   
\vtop{\leftskip 0pt\parindent 0pt\hsize=\tabrulecol  
\ligne{\hfill simple\hfill}
\aff  { 67} {W} {WBWWWBW} {W}
\aff  { 68} {W} {WBWWWBB} {W}
\aff  { 69} {W} {BBWWWBW} {W}
}\hskip 10pt
\vtop{\leftskip 0pt\parindent 0pt\hsize=\tabrulecol  
\ligne{\hfill double\hfill}
\aff  { 70} {W} {BBWWWBB} {W}
}\hskip 10pt
\vtop{\leftskip 0pt\parindent 0pt\hsize=\tabrulecol  
\ligne{\hfill simple\hfill}
\raff { 71} {W} {WWBBBWB} {B}
}\hskip 10pt
\vtop{\leftskip 0pt\parindent 0pt\hsize=\tabrulecol  
\ligne{\hfill double\hfill}
\aff  { 72} {B} {WWBBBWB} {B}
}
\hfill}
\vskip 8pt
\ligne{doubler:\hskip 20pt   
\vtop{\leftskip 0pt\parindent 0pt\hsize=\tabrulecol  
\aff  { 73} {B} {WWBWBWW} {B}
\aff  { 74} {W} {BBBWBWB} {W}
\aff  { 75} {B} {WBWWWBW} {B}
}\hfill fork:\hskip 20pt
\vtop{\leftskip 0pt\parindent 0pt\hsize=\tabrulecol  
\aff  { 76} {W} {WWWWBWB} {W}
\aff  { 77} {W} {WBWBWWW} {W}
}
\hfill} 
\vskip 9pt
\trfn
\vskip 8pt
}
\hfill}

Table~\ref{efx} shows the instructions applied during
the crossing of a simple locomotive
through the fixed switch from the left and from the right, for the cells
of the tracks only. We can notice that the rules involved in the table are
those of the motion of the locomotive on the tracks. The information of those
tables is completed by that of Table~\ref{rfxwit} which shows the rules applied
at the cell~1(1) which has a view on each track arriving 
to the central cell. For that latter table, we again used the colours distinguishing the
front from the rear in a double locomotive.
\vskip 10pt
\ligne{\hfill
\vtop{\leftskip 0pt\parindent 0pt\hsize=300pt
\begin{tab}\label{rfxwit}\leurre
Rules for the cell~$1(1)$ which witnesses the motion on the tracks.
\end{tab}
\vskip-2pt 
\trep
\vskip 8pt
\ligne{
\hfill
\vtop{\leftskip 0pt\parindent 0pt\hsize=\tabrulecol
\ligne{\hfill simple, left\hfill}
\aff  {  2} {B} {WWWWWWW} {B}
\aff  { 44} {B} {WWWWW\Bb BW} {B}
\aff  { 23} {B} {WWWWWW\Bb B} {B}
\aff  {  2} {B} {WWWWWWW} {B}
}\hskip 10pt
\vtop{\leftskip 0pt\parindent 0pt\hsize=\tabrulecol
\ligne{\hfill double, left\hfill}
\aff  {  2} {B} {WWWWWWW} {B}
\aff  { 44} {B} {WWWWW\Bb BW} {B}
\aff  { 62} {B} {WWWWW\Oo B\Bb B} {B}
\aff  { 23} {B} {WWWWWW\Oo B} {B}
\aff  {  2} {B} {WWWWWWW} {B}
}\hskip 10pt
\vtop{\leftskip 0pt\parindent 0pt\hsize=\tabrulecol
\ligne{\hfill simple, right\hfill}
\aff  {  2} {B} {WWWWWWW} {B}
\aff  { 15} {B} {WW\Bb BWWWW} {B}
\aff  { 22} {B} {W\Bb BWWWWW} {B}
\aff  { 19} {B} {\Bb BWWWWWW} {B}
\aff  { 23} {B} {WWWWWW\Bb B} {B}
\aff  {  2} {B} {WWWWWWW} {B}
}\hskip 10pt
\vtop{\leftskip 0pt\parindent 0pt\hsize=\tabrulecol
\ligne{\hfill double, right\hfill}
\aff  {  2} {B} {WWWWWWW} {B}
\aff  { 15} {B} {WW\Bb BWWWW} {B}
\aff  { 46} {B} {W\Bb B\Oo BWWWW} {B}
\aff  { 49} {B} {\Bb B\Oo BWWWWW} {B}
\aff  { 50} {B} {\Oo BWWWWW\Bb B} {B}
\aff  { 23} {B} {WWWWWW\Oo B} {B}
\aff  {  2} {B} {WWWWWWW} {B}
}
\hfill}
\vskip 9pt
\trfn
\vskip 8pt
}
\hfill}

Table~\ref{efx} shows us the rules applied to the elements of the 
tracks traversed by the locomotive when it crosses the switch. Table~\ref{efx}
deals with the     locomotive, whether it is simple or double.
For each side of the switch, the table indicates the group of cells involved by
the arrival of the locomotive to the central cell.

We can see that the rules applied to the central cell have already be seen in
Subsection~\ref{srtrack}. The rules for a simple locomotive coming from the left
are those of the cell~10(4) in Table~\ref{evms}. For a double locomotive from the
left, the rules appear in Tables~\ref{rmotiond} and~\ref{mvdflex} 
for cells whose neighbourhood is \hbox{\tt 2, 5, 7}.  For a simple locomotive
coming from the right, the front rule used here is in Table \ref{mvflex} with 
the neighbourhood \hbox{\tt 2, 5, 7}. When a double locomotive comes from the right, 
Tables~\ref{rmotiond} and~\ref{mvdflex} indicate the corresponding rules.
Figure~\ref{ffx} illustrates the motion of the locomotive: simple or double, whichever
the side from which it arrives to the switch.

\vtop{
\ligne{\hfill
\includegraphics[scale=0.55]{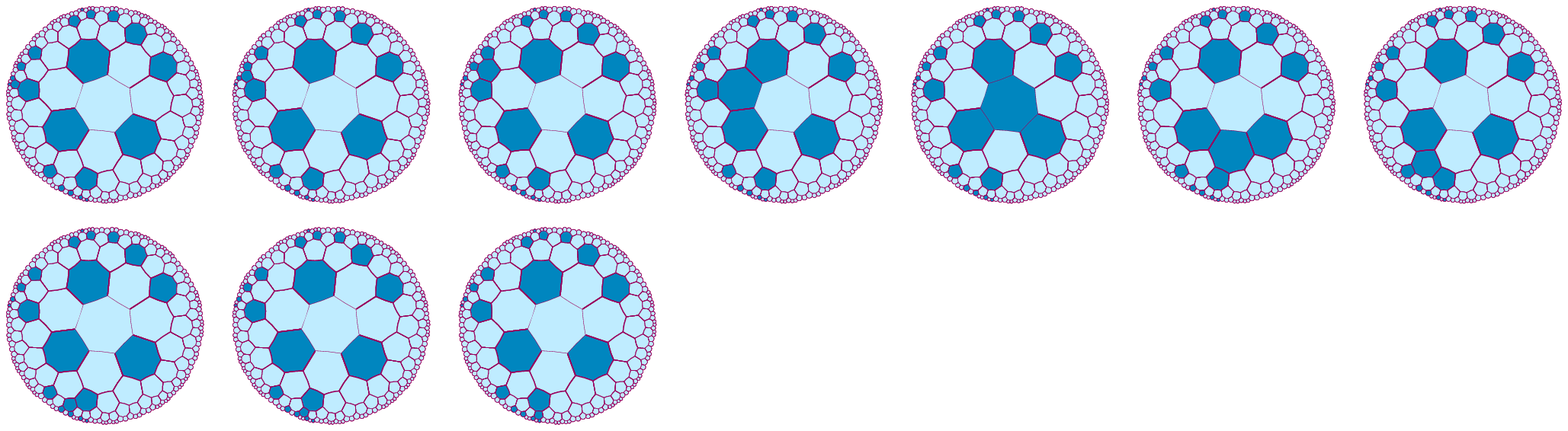}
\hfill}
\vspace{-15pt}
\ligne{\hfill
\includegraphics[scale=0.55]{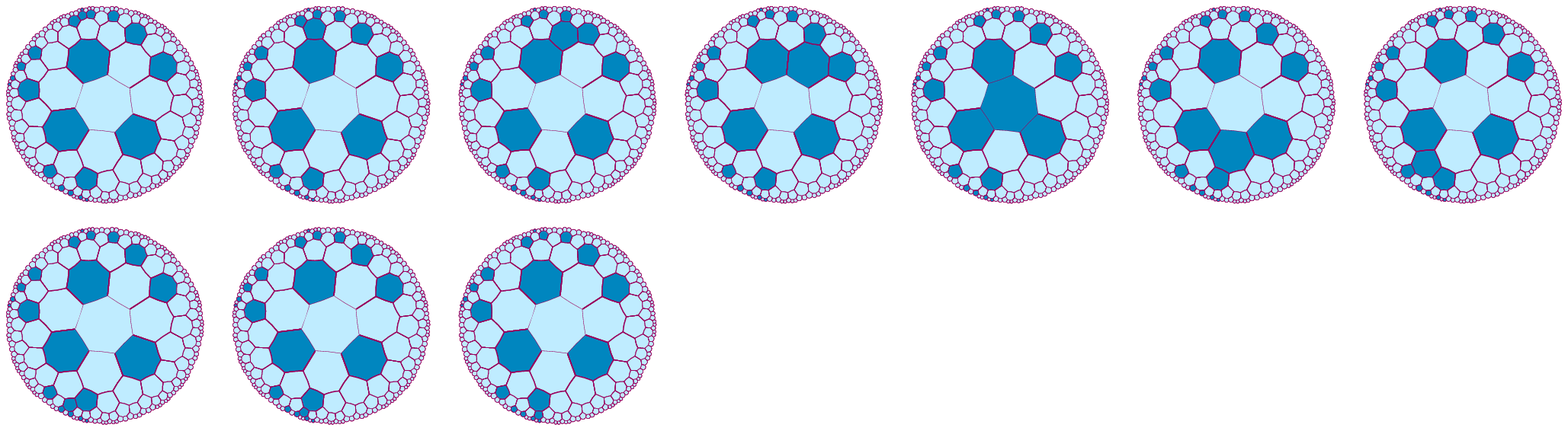}
\hfill}
\vspace{-15pt}
\ligne{\hfill
\includegraphics[scale=0.55]{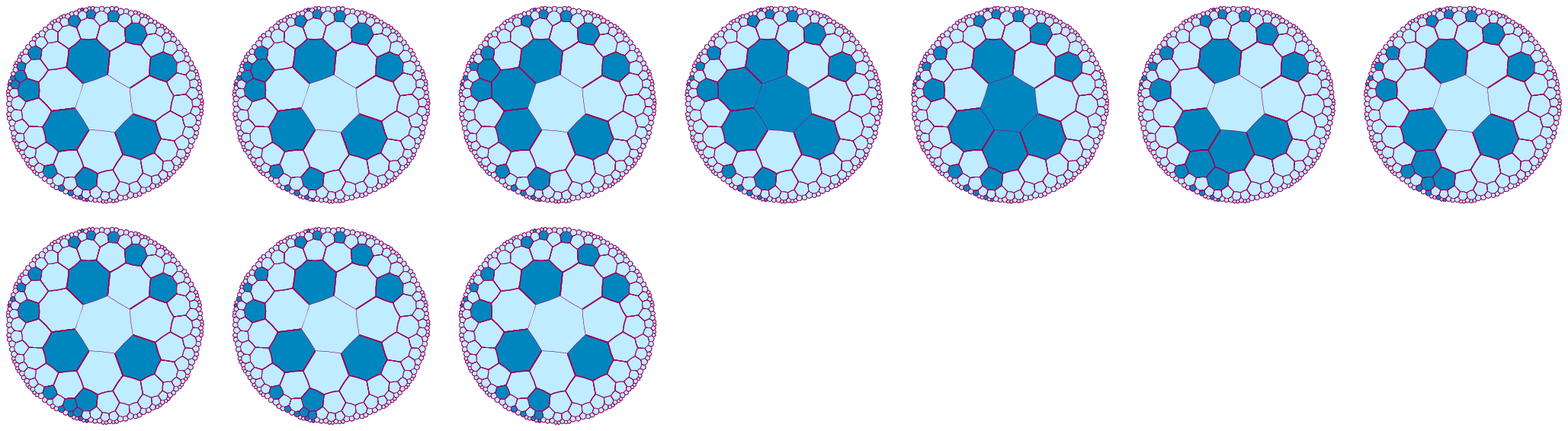}
\hfill}
\vspace{-15pt}
\ligne{\hfill
\includegraphics[scale=0.55]{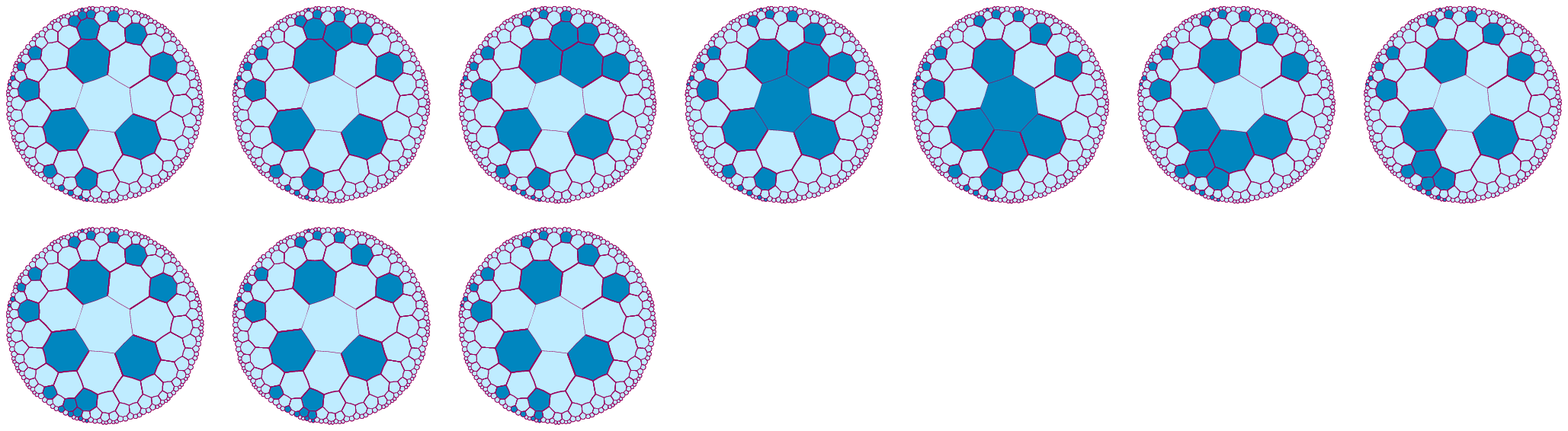}
\hfill}
\ligne{\hfill
\vtop{\leftskip 0pt\parindent 0pt\hsize=300pt
\begin{fig}\label{ffx}
\leurre
Illustration of the motion of a locomotive through a fixed switch.
\end{fig}
}
\hfill}
}

\ligne{\hfill
\vtop{\hsize=340pt
\begin{tab}\label{efx}\leurre
Execution of the rules for the fixed switch when a locomotive
crosses the switch: upper half, simple locomotive; 
lower half, double locomotive.
\end{tab}
\vskip-2pt
\trep
\vskip 8pt
\ligne{\hfill simple locomotive\hfill}
\vskip 5pt
\ligne{\hfill\hbox to 105pt{\hfill\hskip 15pt from the left\hfill}
\hfill\hbox to 85pt{\hfill from the right\hfill}
\hfill\hbox to  110pt{\hfill leaving the cell\hfill}\hfill}
\ligne{\hfill
\vtop{\leftskip 0pt\parindent 0pt\hsize=105pt
\ligne{\hfill\HH{}
\HH{{10$_2$}}\HH{{3$_2$} }\HH{{1$_2$} }\HH{{0$_0$} }\hfill}
\ligne{\hfill\HH{1} \HH{\Rr{24}}\HH{\Rr{38}}\HH{7}\HH{7}\hfill}
\ligne{\hfill\HH{2} \HH{29}\HH{\Rr{40}}\HH{\Rr{45}}\HH{7}\hfill}
\ligne{\hfill\HH{3} \HH{7}\HH{43}\HH{\Rr{24}}\HH{\Rr{16}}\hfill}
\ligne{\hfill\HH{4} \HH{7}\HH{3}\HH{29}\HH{\Rr{24}}\hfill}
\ligne{\hfill\HH{5} \HH{7}\HH{3}\HH{7}\HH{29}\hfill}
\ligne{\hfill\HH{6} \HH{7}\HH{3}\HH{7}\HH{7}\hfill}
\ligne{\hfill\HH{7} \HH{7}\HH{3}\HH{7}\HH{7}\hfill}
\ligne{\hfill\HH{8} \HH{7}\HH{3}\HH{7}\HH{7}\hfill}
}
\hfill
\vtop{\leftskip 0pt\parindent 0pt\hsize=85pt
\ligne{\hfill
\HH{{3$_1$} }\HH{{2$_1$} }\HH{{1$_7$} }\HH{{0$_0$} }\hfill}
\ligne{\hfill \HH{\Rr{40}}\HH{\Rr{41}}\HH{3}\HH{7}\hfill}
\ligne{\hfill \HH{43}\HH{\Rr{17}}\HH{\Rr{38}}\HH{7}\hfill}
\ligne{\hfill \HH{3}\HH{25}\HH{\Rr{40}}\HH{\Rr{45}}\hfill}
\ligne{\hfill \HH{3}\HH{4}\HH{43}\HH{\Rr{24}}\hfill}
\ligne{\hfill \HH{3}\HH{4}\HH{3}\HH{29}\hfill}
\ligne{\hfill \HH{3}\HH{4}\HH{3}\HH{7}\hfill}
\ligne{\hfill \HH{3}\HH{4}\HH{3}\HH{7}\hfill}
\ligne{\hfill \HH{3}\HH{4}\HH{3}\HH{7}\hfill}
}
\hfill
\vtop{\leftskip 0pt\parindent 0pt\hsize=110pt
\ligne{\hfill
\HH{{1$_4$} }\HH{{2$_4$} }\HH{{7$_4$} }\HH{{8$_4$} }\HH{{9$_4$} } \hfill}
\ligne{\hfill\HH{7}\HH{14}\HH{7}\HH{4}\HH{7}\hfill}
\ligne{\hfill\HH{7}\HH{14}\HH{7}\HH{4}\HH{7}\hfill}
\ligne{\hfill\HH{7}\HH{14}\HH{7}\HH{4}\HH{7}\hfill}
\ligne{\hfill\HH{\Rr{16}}\HH{14}\HH{7}\HH{4}\HH{7}\hfill}
\ligne{\hfill\HH{\Rr{24}}\HH{\Rr{63}}\HH{7}\HH{4}\HH{7}\hfill}
\ligne{\hfill\HH{29}\HH{\Rr{31}}\HH{\Rr{32}}\HH{4}\HH{7}\hfill}
\ligne{\hfill\HH{7}\HH{34}\HH{\Rr{24}}\HH{\Rr{36}}\HH{7}\hfill}
\ligne{\hfill\HH{7}\HH{14}\HH{29}\HH{\Rr{17}}\HH{\Rr{16}}\hfill}
}
\hfill}
\vskip 8pt
\trfn
\vskip 8pt
\ligne{\hfill double locomotive\hfill}
\vskip 5pt
\ligne{\hfill\hbox to 105pt{\hfill\hskip 15pt from the left\hfill}
\hfill\hbox to 85pt{\hfill from the right\hfill}
\hfill\hbox to  110pt{\hfill leaving the cell\hfill}\hfill}
\ligne{\hfill
\vtop{\leftskip 0pt\parindent 0pt\hsize=105pt
\ligne{\hfill\HH{}
\HH{{10$_2$}}\HH{{3$_2$} }\HH{{1$_2$} }\HH{{0$_0$} }\hfill}
\ligne{\hfill\HH{1} \HH{\Rr{51}}\HH{59}\HH{\Rr{45}}\HH{7}\hfill}
\ligne{\hfill\HH{2} \HH{29}\HH{\Rr{60}}\HH{66}\HH{\Rr{16}}\hfill}
\ligne{\hfill\HH{3} \HH{7}\HH{43}\HH{\Rr{51}}\HH{47}\hfill}
\ligne{\hfill\HH{4} \HH{7}\HH{3}\HH{29}\HH{\Rr{51}}\hfill}
\ligne{\hfill\HH{5} \HH{7}\HH{3}\HH{7}\HH{29}\hfill}
\ligne{\hfill\HH{6} \HH{7}\HH{3}\HH{7}\HH{7}\hfill}
\ligne{\hfill\HH{7} \HH{7}\HH{3}\HH{7}\HH{7}\hfill}
\ligne{\hfill\HH{8} \HH{7}\HH{3}\HH{7}\HH{7}\hfill}
}\hfill
\vtop{\leftskip 0pt\parindent 0pt\hsize=85pt
\ligne{\hfill
\HH{{3$_1$} }\HH{{2$_1$} }\HH{{1$_7$} }\HH{{0$_0$} }\hfill}
\ligne{\hfill\HH{\Rr{60}}\HH{61}\HH{\Rr{38}}\HH{7}\hfill}
\ligne{\hfill\HH{43}\HH{\Rr{48}}\HH{59}\HH{\Rr{45}}\hfill}
\ligne{\hfill\HH{3}\HH{25}\HH{\Rr{60}}\HH{66}\hfill}
\ligne{\hfill\HH{3}\HH{4}\HH{43}\HH{\Rr{51}}\hfill}
\ligne{\hfill\HH{3}\HH{4}\HH{3}\HH{29}\hfill}
\ligne{\hfill\HH{3}\HH{4}\HH{3}\HH{7}\hfill}
\ligne{\hfill\HH{3}\HH{4}\HH{3}\HH{7}\hfill}
\ligne{\hfill\HH{3}\HH{4}\HH{3}\HH{7}\hfill}
}\hfill
\vtop{\leftskip 0pt\parindent 0pt\hsize=110pt
\ligne{\hfill
\HH{{1$_4$} }\HH{{2$_4$} }\HH{{7$_4$} }\HH{{8$_4$} }\HH{{9$_4$} }\hfill}
\ligne{\hfill \HH{7}\HH{14}\HH{7}\HH{4}\HH{7}\hfill}
\ligne{\hfill \HH{7}\HH{14}\HH{7}\HH{4}\HH{7}\hfill}
\ligne{\hfill \HH{\Rr{16}}\HH{14}\HH{7}\HH{4}\HH{7}\hfill}
\ligne{\hfill \HH{47}\HH{\Rr{63}}\HH{7}\HH{4}\HH{7}\hfill}
\ligne{\hfill \HH{\Rr{51}}\HH{65}\HH{\Rr{32}}\HH{4}\HH{7}\hfill}
\ligne{\hfill \HH{29}\HH{\Rr{53}}\HH{54}\HH{\Rr{36}}\HH{7}\hfill}
\ligne{\hfill \HH{7}\HH{34}\HH{\Rr{51}}\HH{56}\HH{\Rr{16}}\hfill}
\ligne{\hfill \HH{7}\HH{14}\HH{29}\HH{\Rr{48}}\HH{47}\hfill}
}
\hfill}
\vskip 9pt
\trfn
\vskip 8pt
}
\hfill}

Figure~\ref{ffx} illustrates the four motions we have to consider for the
fixed switch for which Tables~\ref{efx} gives the rules
used for such motions.

\subsubsection{The rules for the doubler and for the fork}\label{srdblfrk}

As clear from Table~\ref{rfixed}, a few rules only are needed by the doubler and by 
the fork. As the doubler contains both the fork and the fixed switch, 
Table~\ref{rfixed} displays the three additional rules required by the doubler
before the two ones required by the fork as tested in the configuration of 
the left-hand side picture of Figure~\ref{stab_dblfrk}. Here, we distinguish between
the two locomotives created by the cell~4(1) by giving them colours: green for the
locomotive which will follow the green path, dark pink for the one which will go along the 
pink path.

In Table~\ref{rdblwit}, we give the rules used by the doubler when the locomotives 
cross the cells 4(1), 1(1) and~0(0). Note that when the locomotive enters the cell~4(1), 
at the next time, two locomotives leave the cell, which is witnessed by the cell, see
rule~74 in Table~\ref{rdblwit}. The cell~1(1) also witnesses the duplication
as shown by rules~37, 73, 75 and~50. That last rule witnesses the junction of
the two simple locomotives into a double one. Note that the cell~4(1) is applied a 
sequence of rules which differ from a sequence indicated in Subsection~\ref{srtrack} 
by the witness rule: instead of rule~29 as in the motion rules, we have here rule~74 as 
the cell can see two locomotives created in its neighbours~1 and~3. In the cell~0(0), 
rule~66 witnesses the junction of the two simple locomotives into a double one. This is 
why in the sequence of rules~7, 16, 47, 51 and~29 in the crossing of a cell by a double 
locomotive, see the rules for the cell~1(6) in Table~\ref{evdd}, rule~47 is replaced by 
rule~66 as the occurrence of the second cell, the rear, appears from the side which is
opposite to the expected one.

\ligne{\hfill  
\vtop{\leftskip 0pt\parindent 0pt\hsize=240pt 
\begin{tab}\label{rdblwit}\leurre
Rules for the cells~$4(1)$, $0(0)$ and $1(1)$ which witness the motion of the 
locomotives in the doubler.
\end{tab}
\vskip-2pt 
\trep
\vskip 8pt
\ligne{\hfill   
\vtop{\leftskip 0pt\parindent 0pt\hsize=\tabrulecol
\ligne{4(1):\hfill} 
\aff  {  7} {W} {WBWWBWB} {W} 
\raff { 16} {W} {WBWWB\Bb BB} {B} 
\raff { 24} {B} {WBWWBWB} {W}
\aff  { 74} {W} {\Gg BB\Pp BWBWB} {W}
\aff  {  7} {W} {WBWWBWB} {W}
}\hskip 20pt
\vtop{\leftskip 0pt\parindent 0pt\hsize=\tabrulecol
\ligne{0(0):\hfill} 
\aff  {  7} {W} {WBWWBWB} {W}
\raff { 16} {W} {WBWWB\Gg BB} {B} 
\aff  { 66} {\Gg B} {WBW\Pp BBWB} {B}
\raff { 51} {\Pp B} {\Gg BBWWBWB} {W}
\aff  { 29} {W} {\Pp BBWWBWB} {W}
\aff  {  7} {W} {WBWWBWB} {W}
}\hskip 20pt
\vtop{\leftskip 0pt\parindent 0pt\hsize=\tabrulecol
\ligne{1(1):\hfill} 
\aff  {  2} {B} {WWWWWWW} {B}
\aff  { 37} {B} {WWW\Bb BWWW} {B}
\aff  { 73} {B} {WW\Pp BW\Gg BWW} {B}
\aff  { 75} {B} {W\Pp BWWW\Gg BW} {B}
\aff  { 50} {B} {\Pp BWWWWW\Gg B} {B}
\aff  {  2} {B} {WWWWWWW} {B}
}
\hfill}  
\vskip 9pt
\trfn
\vskip 8pt
}
\hfill}

\ligne{\hfill  
\vtop{\hsize=280pt 
\begin{tab}\label{edbl}\leurre
Upper, lower part: execution of the rules for the doubler, the fork respectively,
corresponding to the illustrations of Figure~{\rm\ref{stab_dblfrk}}.
\end{tab}
\vskip-2pt
\trep
\vskip 8pt
\ligne{\hfill doubler\hfill}
\vskip 5pt
\ligne{\hfill\HH{}
\HH{{12$_4$}}\HH{{4$_1$} }\HH{{2$_2$} }\HH{{1$_2$} }\HH{{0$_0$} }\HH{{1$_7$} }\HH{{2$_1$} }\HH{{3$_1$} }\HH{{1$_4$} }\HH{{2$_4$} }\HH{{7$_4$} }\HH{{8$_4$} }
\hfill}
\ligne{\hfill\HH{1}
\HH{25}\HH{\Rr{24}}\HH{\Rr{36}}\HH{4}\HH{7}\HH{3}\HH{7}\HH{\Rr{41}}\HH{7}\HH{14}\HH{7}\HH{4}\hfill}
\ligne{\hfill\HH{2}
\HH{4}\HH{74}\HH{\Rr{17}}\HH{\Rr{36}}\HH{7}\HH{3}\HH{\Rr{16}}\HH{\Rr{17}}\HH{7}\HH{14}\HH{7}\HH{4}\hfill}
\ligne{\hfill\HH{3}
\HH{4}\HH{7}\HH{25}\HH{\Rr{17}}\HH{\Rr{16}}\HH{\Rr{38}}\HH{\Rr{24}}\HH{25}\HH{7}\HH{14}\HH{7}\HH{4}\hfill}
\ligne{\hfill\HH{4}
\HH{4}\HH{7}\HH{4}\HH{25}\HH{66}\HH{\Rr{60}}\HH{29}\HH{4}\HH{\Rr{16}}\HH{14}\HH{7}\HH{4}\hfill}
\ligne{\hfill\HH{5}
\HH{4}\HH{7}\HH{4}\HH{25}\HH{\Rr{51}}\HH{43}\HH{7}\HH{4}\HH{47}\HH{\Rr{63}}\HH{7}\HH{4}\hfill}
\ligne{\hfill\HH{6}
\HH{4}\HH{7}\HH{4}\HH{4}\HH{29}\HH{3}\HH{7}\HH{4}\HH{\Rr{51}}\HH{65}\HH{\Rr{32}}\HH{4}\hfill}
\ligne{\hfill\HH{7}
\HH{4}\HH{7}\HH{4}\HH{4}\HH{7}\HH{3}\HH{7}\HH{4}\HH{29}\HH{\Rr{53}}\HH{54}\HH{\Rr{36}}\hfill}
\ligne{\hfill\HH{8}
\HH{4}\HH{7}\HH{4}\HH{4}\HH{7}\HH{3}\HH{7}\HH{4}\HH{7}\HH{34}\HH{\Rr{51}}\HH{56}\hfill}
\vskip 9pt
\trfn
\vskip 8pt
\ligne{\hfill fork\hfill}
\vskip 5pt
\ligne{\hfill\HH{}
\HH{{10$_1$}}\HH{{4$_1$} }\HH{{1$_1$} }\HH{{1$_2$} }\HH{{1$_3$} }\HH{{3$_3$} }\HH{{7$_3$} }\HH{{1$_7$} }\HH{{1$_6$} }\HH{{4$_6$} }\HH{{5$_7$} }
\hfill}
\ligne{\hfill\HH{1}
\HH{29}\HH{\Rr{17}}\HH{\Rr{16}}\HH{4}\HH{7}\HH{7}\HH{4}\HH{4}\HH{4}\HH{4}\HH{7}\hfill}
\ligne{\hfill\HH{2}
\HH{7}\HH{25}\HH{\Rr{24}}\HH{\Rr{36}}\HH{7}\HH{7}\HH{4}\HH{\Rr{41}}\HH{4}\HH{4}\HH{7}\hfill}
\ligne{\hfill\HH{3}
\HH{7}\HH{4}\HH{74}\HH{\Rr{17}}\HH{\Rr{16}}\HH{7}\HH{4}\HH{\Rr{17}}\HH{\Rr{36}}\HH{4}\HH{7}\hfill}
\ligne{\hfill\HH{4}
\HH{7}\HH{4}\HH{7}\HH{25}\HH{\Rr{24}}\HH{\Rr{16}}\HH{4}\HH{25}\HH{\Rr{17}}\HH{\Rr{36}}\HH{7}\hfill}
\ligne{\hfill\HH{5}
\HH{7}\HH{4}\HH{7}\HH{4}\HH{29}\HH{\Rr{24}}\HH{\Rr{36}}\HH{4}\HH{25}\HH{\Rr{17}}\HH{\Rr{16}}\hfill}
\vskip 9pt
\trfn
\vskip 8pt
}
\hfill}

\vtop{
\ligne{\hfill
\includegraphics[scale=0.55]{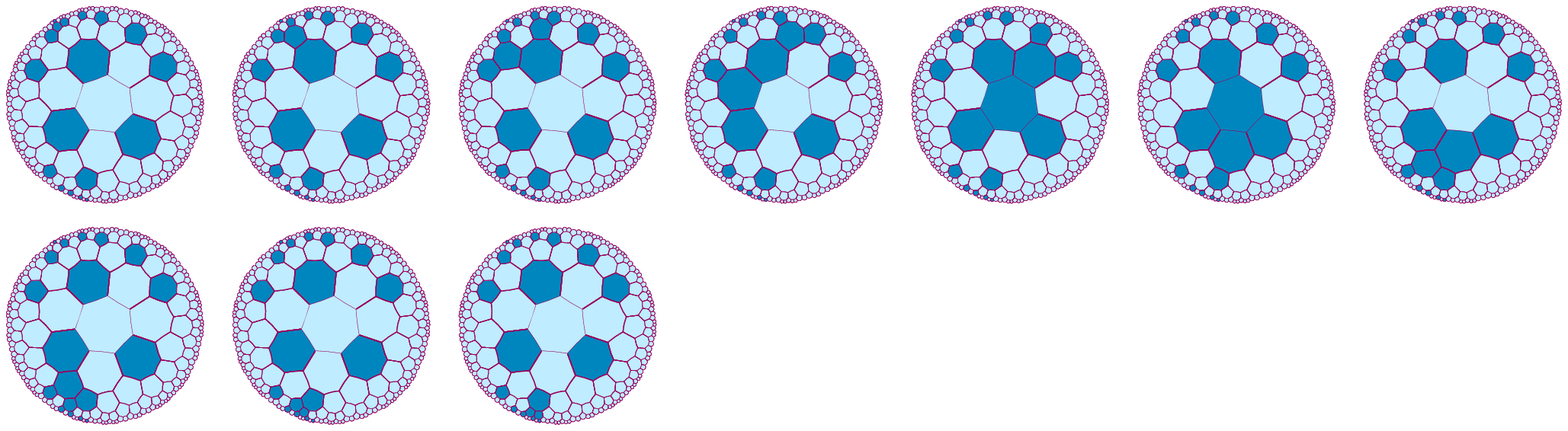}
\hfill}
\vspace{-15pt}
\ligne{\hfill
\includegraphics[scale=0.55]{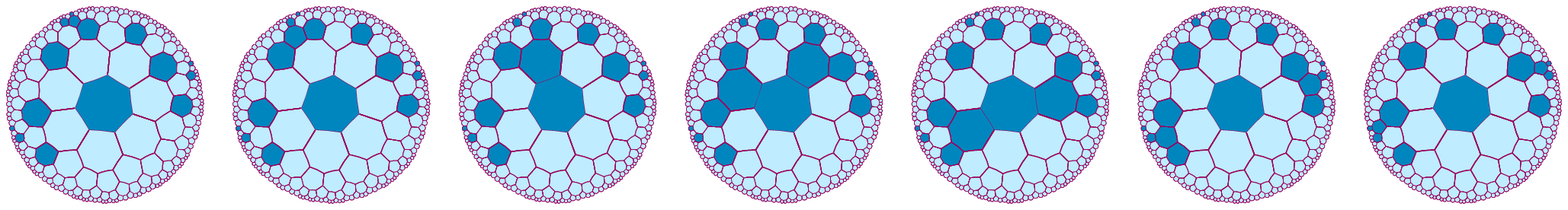}
\hfill}
\begin{fig}\label{fdblfrk}
Illustration of the crossing of the doubler, above, and of the fork, below, by the
locomotive.
\end{fig}
}

\subsection{The rules for the selector}\label{srsel}

The selector is the last structure we need to implement round-abouts. The new rules needed
by the structure are given in Table~\ref{rsel} while the execution of the rules used
by the crossing of a locomotive are given in Table~\ref{esel}: the left-, right-hand side 
sub-table gives the rules used by a simple, double locomotive respectively.

\ligne{\hfill 
\vtop{\leftskip 0pt\parindent 0pt\hsize=\tabruleli  
\begin{tab}\label{rsel}
\leurre
Rules for the locomotive through the selector.
\end{tab}
\vskip-2pt
\trep
\vskip 8pt
\ligne{\hfill simple locomotive\hfill}
\vskip 0pt
\ligne{\hfill   
\vtop{\leftskip 0pt\parindent 0pt\hsize=\tabrulecol  
\aff  { 78} {W} {BWBBWBW} {W}
\aff  { 79} {W} {BBWBBBW} {W}
\aff  { 80} {B} {WBBWWBB} {B}
\aff  { 81} {B} {BBWBWWW} {B}
\aff  { 82} {B} {BBWWBWW} {B}
\raff { 83} {W} {BBWBBBB} {B}
\aff  { 84} {B} {BBWWWBW} {B}
}\hskip 10pt
\vtop{\leftskip 0pt\parindent 0pt\hsize=\tabrulecol  
\aff  { 85} {B} {BWBWWWB} {B}
\raff { 86} {W} {BWBBWBB} {B}
\aff  { 87} {B} {WWWWBBB} {B}
\raff { 88} {B} {BBWBBBW} {W}
\aff  { 89} {B} {BBBWWBB} {B}
\aff  { 90} {B} {WBBBWWW} {B}
\aff  { 91} {B} {BBWWWWB} {B}
}\hskip 10pt
\vtop{\leftskip 0pt\parindent 0pt\hsize=\tabrulecol  
\raff { 92} {B} {BWBBWBW} {W}
\raff { 93} {B} {BWWWBBW} {W}
\aff  { 94} {W} {BBBBBBW} {W}
\aff  { 95} {B} {BWBBWWW} {B}
\aff  { 96} {W} {BBBBBWW} {W}
\raff { 97} {B} {WWWBWWB} {W}
\raff { 98} {W} {WBWWBBW} {B}
}\hskip 10pt
\vtop{\leftskip 0pt\parindent 0pt\hsize=\tabrulecol  
\aff  { 99} {W} {BBWWBBW} {W}
\aff  {100} {B} {WWBWWBB} {B}
\aff  {101} {B} {WWBBWWB} {B}
\aff  {102} {B} {WBWWWWB} {B}
\aff  {103} {B} {WWBBWBW} {B}
\aff  {104} {W} {WBBBWWW} {W}
}
\hfill} 
\vskip 9pt
\ligne{\hfill double locomotive\hfill}
\vskip 0pt
\ligne{\hfill 
\vtop{\leftskip 0pt\parindent 0pt\hsize=\tabrulecol  
\aff  {105} {B} {BBWBBWW} {B}
\aff  {106} {B} {BBWWBBW} {B}
\aff  {107} {B} {BBWBBBB} {B}
\aff  {108} {B} {BBWWWBB} {B}
}\hskip 10pt
\vtop{\leftskip 0pt\parindent 0pt\hsize=\tabrulecol  
\aff  {109} {B} {BBBWWWB} {B}
\raff {110} {B} {BWBBWBB} {W}
\aff  {111} {B} {BWWWBBB} {B}
\raff {112} {B} {BBBBBBW} {W}
}\hskip 10pt
\vtop{\leftskip 0pt\parindent 0pt\hsize=\tabrulecol  
\raff {113} {B} {BBBBWWW} {W}
\aff  {114} {W} {WBBBBBW} {W}
\raff {115} {B} {WBWWBWW} {W}
\aff  {116} {B} {WBWWBBW} {B}
}\hskip 10pt
\vtop{\leftskip 0pt\parindent 0pt\hsize=\tabrulecol  
\raff {117} {W} {WWBBWWB} {B}
\aff  {118} {W} {BWWBBBW} {W}
\aff  {119} {B} {BWWWBWW} {B}
\aff  {120} {B} {WWBWBBW} {B}
}
\hfill} 
\vskip 9pt
\trfn
\vskip 8pt
} 
\hfill}

In both sub-tables of Table~\ref{esel}, we can see that the track leading the locomotive
to the selector make use of motion rules examined in Sub-section~\ref{srtrack},
the cell 1(6) excepted. That cell, which constitutes the entrance to the selector
has a specific neighbourhood involving five milestones.

\ligne{\hfill
\vtop{\hsize=260pt
\begin{tab}\label{esel}\leurre
Execution of the rules for a locomotive passing through the selector.
\end{tab}
\vskip-2pt
\trep
\vskip 8pt
\ligne{\hfill simple locomotive\hfill}
\vskip 8pt
\ligne{\hfill\HH{}
\HH{{9$_6$} }\HH{{10$_6$}}\HH{{4$_6$} }\HH{{1$_6$} }\HH{{0$_0$} }\HH{{1$_1$} }\HH{{2$_1$} }\HH{{7$_1$} }\HH{{1$_7$} }\HH{{1$_5$} }
\hfill}
\ligne{\hfill\HH{1}
\HH{29}\HH{\Rr{17}}\HH{\Rr{36}}\HH{79}\HH{78}\HH{7}\HH{4}\HH{7}\HH{57}\HH{64}\hfill}
\ligne{\hfill\HH{2}
\HH{7}\HH{25}\HH{\Rr{17}}\HH{\Rr{83}}\HH{78}\HH{7}\HH{4}\HH{7}\HH{57}\HH{64}\hfill}
\ligne{\hfill\HH{3}
\HH{7}\HH{4}\HH{25}\HH{\Rr{88}}\HH{\Rr{86}}\HH{7}\HH{4}\HH{7}\HH{90}\HH{87}\hfill}
\ligne{\hfill\HH{4}
\HH{7}\HH{4}\HH{4}\HH{94}\HH{\Rr{92}}\HH{\Rr{16}}\HH{4}\HH{7}\HH{95}\HH{\Rr{93}}\hfill}
\ligne{\hfill\HH{5}
\HH{7}\HH{4}\HH{4}\HH{99}\HH{96}\HH{\Rr{24}}\HH{\Rr{36}}\HH{7}\HH{101}\HH{\Rr{98}}\hfill}
\ligne{\hfill\HH{6}
\HH{7}\HH{4}\HH{4}\HH{79}\HH{78}\HH{29}\HH{\Rr{17}}\HH{\Rr{16}}\HH{103}\HH{64}\hfill}
\ligne{\hfill\HH{7}
\HH{7}\HH{4}\HH{4}\HH{79}\HH{78}\HH{7}\HH{25}\HH{\Rr{24}}\HH{57}\HH{64}\hfill}
\vskip 9pt
\trfn
%
\vskip 8pt
\ligne{\hfill double locomotive\hfill}
\vskip 8pt
\ligne{\hfill\HH{}
\HH{{9$_6$} }\HH{{10$_6$}}\HH{{4$_6$} }\HH{{1$_6$} }\HH{{0$_0$} }\HH{{1$_4$} }\HH{{2$_5$} }\HH{{5$_5$} }\HH{{12$_4$}}\HH{{1$_7$} }\HH{{1$_5$} }
\hfill}
\ligne{\hfill\HH{1}
\HH{29}\HH{\Rr{48}}\HH{56}\HH{\Rr{83}}\HH{78}\HH{4}\HH{7}\HH{7}\HH{4}\HH{57}\HH{64}\hfill}
\ligne{\hfill\HH{2}
\HH{7}\HH{25}\HH{\Rr{48}}\HH{107}\HH{\Rr{86}}\HH{4}\HH{7}\HH{7}\HH{4}\HH{90}\HH{87}\hfill}
\ligne{\hfill\HH{3}
\HH{7}\HH{4}\HH{25}\HH{\Rr{112}}\HH{\Rr{110}}\HH{\Rr{36}}\HH{7}\HH{7}\HH{4}\HH{\Rr{113}}\HH{111}\hfill}
\ligne{\hfill\HH{4}
\HH{7}\HH{4}\HH{4}\HH{118}\HH{114}\HH{\Rr{17}}\HH{\Rr{16}}\HH{7}\HH{4}\HH{\Rr{117}}\HH{116}\hfill}
\ligne{\hfill\HH{5}
\HH{7}\HH{4}\HH{4}\HH{79}\HH{78}\HH{25}\HH{\Rr{24}}\HH{\Rr{16}}\HH{4}\HH{57}\HH{120}\hfill}
\ligne{\hfill\HH{6}
\HH{7}\HH{4}\HH{4}\HH{79}\HH{78}\HH{4}\HH{29}\HH{\Rr{24}}\HH{\Rr{36}}\HH{57}\HH{64}\hfill}
\ligne{\hfill\HH{7}
\HH{7}\HH{4}\HH{4}\HH{79}\HH{78}\HH{4}\HH{7}\HH{29}\HH{\Rr{17}}\HH{57}\HH{64}\hfill}
\vskip 9pt
\trfn
\vskip 8pt
}
\hfill}

\ligne{\hfill  
\vtop{\leftskip 0pt\parindent 0pt\hsize=340pt 
\begin{tab}\label{rselwit}\leurre
Rules for the cells~$1(7)$ and $1(5)$, $1(6)$ and $0(0)$ which witness the motion of the 
locomotives in the doubler. 
\end{tab}
\vskip-2pt 
\trep
\vskip 8pt
\ligne{\hskip 80pt 1(7)\hfill 1(5)\hskip 80pt}
\ligne{\hfill   
\vtop{\leftskip 0pt\parindent 0pt\hsize=\tabrulecol
\ligne{\hfill simple\hfill} 
\aff  { 57} {B} {WWBBWWW} {B}
\aff  { 90} {B} {W\Bb BBBWWW} {B}
\aff  { 95} {B} {\Bb BWBBWWW} {B}
\aff  {101} {B} {WWBBWW\Pp B} {B}
\aff  {103} {B} {WWBBW\Pp BW} {B}
\aff  { 57} {B} {WWBBWWW} {B}
}\hskip 15pt
\vtop{\leftskip 0pt\parindent 0pt\hsize=\tabrulecol
\ligne{\hfill double\hfill} 
\aff  { 57} {B} {WWBBWWW} {B}
\aff  { 90} {B} {W\Bb BBBWWW} {B}
\raff {113} {B} {\Bb B\Oo BBBWWW} {\Rr W}
\raff {117} {\Rr W} {WWBBWW\Oo B} {B}
\aff  { 57} {B} {WWBBWWW} {B}
}
\hfill
\vtop{\leftskip 0pt\parindent 0pt\hsize=\tabrulecol
\ligne{\hfill simple\hfill} 
\aff  { 64} {B} {WWWWBBW} {B}
\aff  { 87} {B} {WWWWBB\Bb B} {B}
\raff { 93} {B} {\Bb BWWWBBW} {\Rr W}
\raff { 98} {\Rr W} {W\Gg BWWBBW} {B}
\aff  { 64} {B} {WWWWBBW} {B}
}\hskip 15pt
\vtop{\leftskip 0pt\parindent 0pt\hsize=\tabrulecol
\ligne{\hfill double\hfill} 
\aff  { 64} {B} {WWWWBBW} {B}
\aff  { 87} {B} {WWWWBB\Bb B} {B}
\aff  {111} {B} {\Bb BWWWBB\Oo B} {B}
\aff  {116} {B} {W\Gg BWWBBW} {B}
\aff  {120} {B} {WW\Gg BWBBW} {B}
\aff  { 64} {B} {WWWWBBW} {B}
}
\hfill}
\vskip 8pt
\ligne{\hskip 80pt 1(6)\hfill 0(0)\hskip 80pt}
\ligne{\hfill   
\vtop{\leftskip 0pt\parindent 0pt\hsize=\tabrulecol
\ligne{\hfill simple\hfill} 
\aff  { 79} {W} {BBWBBBW} {W}
\raff { 83} {W} {BBWBBB\Bb B} {B}
\raff { 88} {\Bb B} {BBWBBBW} {W}
\aff  { 94} {W} {BB\Bb BBBBW} {W}
\aff  { 99} {W} {BBW\Rr WBBW} {W}
\aff  { 79} {W} {BBWBBBW} {W}
}\hskip 15pt
\vtop{\leftskip 0pt\parindent 0pt\hsize=\tabrulecol
\ligne{\hfill double\hfill} 
\aff  { 79} {W} {BBWBBBW} {W}
\raff { 83} {W} {BBWBBB\Bb B} {B}
\aff  {107} {\Bb B} {BBWBBB\Oo B} {B}
\raff {112} {\Oo B} {BB\Bb BBBBW} {W}
\aff  {118} {W} {B\Rr WWBBBW} {W}
\aff  { 79} {W} {BBWBBBW} {W}
}\hfill
\vtop{\leftskip 0pt\parindent 0pt\hsize=\tabrulecol
\ligne{\hfill simple\hfill} 
\aff  { 78} {W} {BWBBWBW} {W}
\raff { 86} {W} {BWBBWB\Bb B} {B}
\raff { 92} {\Bb B} {BWBBWBW} {W}
\aff  { 96} {W} {B\Oo BBB\Gg B\Rr WW} {W}
\aff  { 78} {W} {BWBBWBW} {W}
}\hskip 15pt
\vtop{\leftskip 0pt\parindent 0pt\hsize=\tabrulecol
\ligne{\hfill double\hfill} 
\aff  { 78} {W} {BWBBWBW} {W}
\raff { 86} {W} {BWBBWB\Bb B} {B}
\raff {110} {\Bb B} {BWBBWB\Oo B} {W}
\aff  {114} {W} {\Rr W\Oo BBB\Gg BBW} {W}
\aff  { 78} {W} {BWBBWBW} {W}
}
\hfill}
\vskip 9pt
\trfn
\vskip 8pt
}
\hfill}

Among them, the cells~1(7)
and~1(5) which constitute the sensors of the selector: they detect whether a simple or
a double locomotive arrived at the cell~0(0). Table~\ref{rselwit} shows the rules
applied at the cells~1(7), 1(5), 1(6) and~0(0). For each cell, the table gives the
rules when a simple locomotive arrives and then, when a double one arrives. Note
that \zz{\Rr W} indicates that the cell~1(7) or 1(5) became white for one time in order
to cancell the locomotive prepared for the corresponding path.

Figure~\ref{fsel} illustrates the motion of the locomotive in the selector, whether
it is simple or double.

At this point, it is important to point at the fact that for the cell~1(6), side~1
is not shared with~0(0) but with a milestone of~1(6), namely 2(7). In the same way,
for~0(0), side~1 is not shared with a cell of the tracks, it is also shared with a
milestone, with the cell~1(7). The reason of these choices lies in the fact that
the idle neighbourhood of 1(6) coincide with the rotated form of a neighbourhood
of~0(0) in the selector: this can be seen with rules~79 and~86 whose neighbourhood are
rotated forms of each other and which, consequently, are rotationally incompatible.

\vtop{
\ligne{\hfill
\includegraphics[scale=0.55]{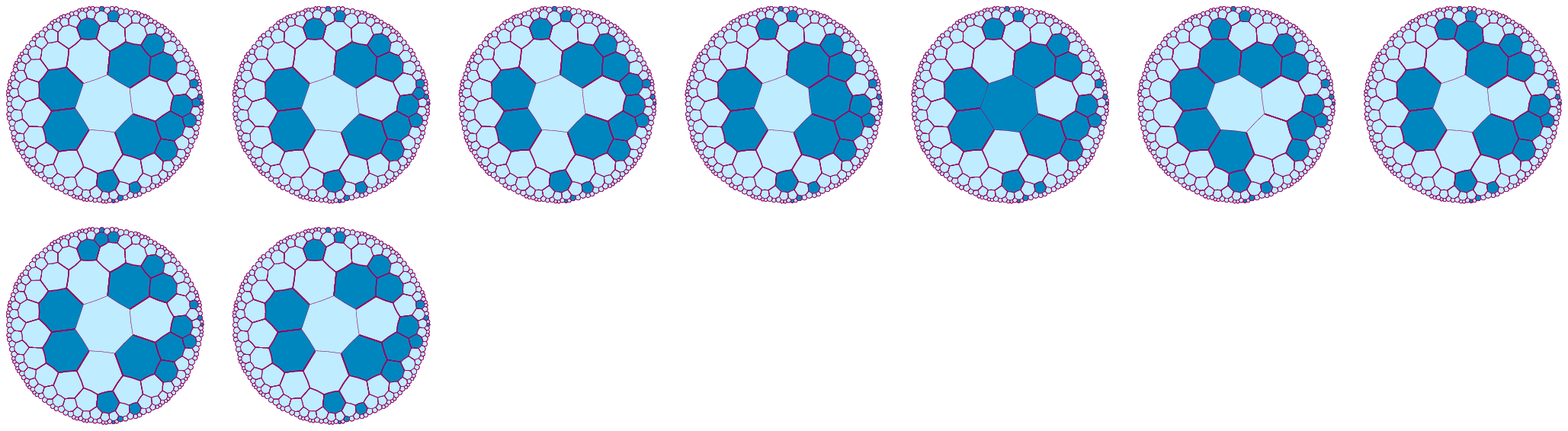}
\hfill}
\vspace{-15pt}
\ligne{\hfill
\includegraphics[scale=0.55]{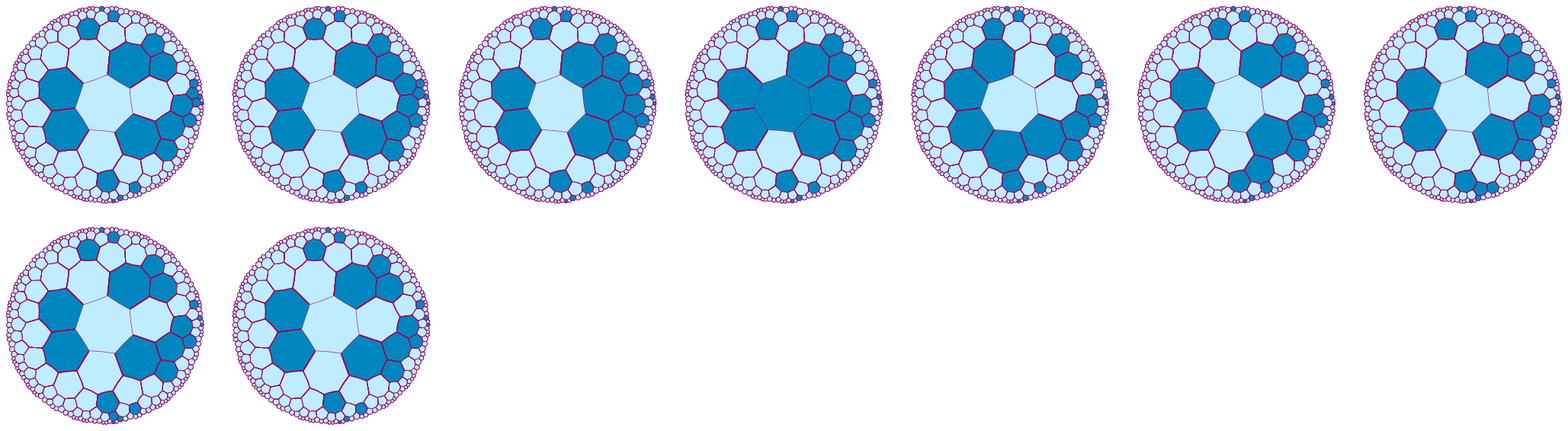}
\hfill}
\begin{fig}\label{fsel}\leurre
Illustration of the crossing of the selector: above, by a simple locomotive, 
below, by a double one.
\end{fig}
}

\subsection{The rules for the controller}\label{srctrl}

Let us now consider the rules for the controller of the active switches. The rules are
displayed by Table~\ref{rctrl}. As mentioned in the table itself, the two columns in 
the left-hand side deal with the passage of the locomotive while the last column deals
with the change of colour of the controller. We remind the reader that the colour
of the controller is the colour of the cell~1(3) in Figure~\ref{stab_ctrl}. 
Table~\ref{ectrlb} and Figure~\ref{rctrl} illustrate the crossing of a black controller 
by the locomotive.
All cells of the track obey the rules we have considered for the tracks for 
three-milestoned rules. As an example, the cell 1(4) is applied the same rules
as the cell 1(6) in Tables~\ref{rmotion} and~\ref{evms}.
We shall look closely at the cell 1(3), the control cell of the structure and at
the central cell 0(0). 

\ligne{\hfill 
\vtop{\leftskip 0pt\parindent 0pt\hsize=\tabrulecol
\advance\hsize by\tabrulecol\advance \hsize by \tabrulecol
\advance\hsize by 40pt  
\begin{tab}\label{rctrl}
\leurre
Rules for the control: passage of the locomotive and signal for changing the selected
track.
\end{tab}
\vskip-2pt
\trep
\vskip 8pt
\ligne{\hfill passage of the locomotive\hfill
\hbox to \tabrulecol{\hfill signal\hfill}}
\vskip 0pt
\ligne{\hfill   
\vtop{\leftskip 0pt\parindent 0pt\hsize=\tabrulecol  
\ligne{\hfill black\hfill}
\aff  {121} {W} {WWWWWBB} {W}
\aff  {122} {W} {BWWBWWW} {W}
\aff  {123} {W} {BWBBWWW} {W}
\aff  {124} {W} {BBWBWWW} {W}
}\hskip 10pt
\vtop{\leftskip 0pt\parindent 0pt\hsize=\tabrulecol  
\ligne{\hfill white\hfill}
\aff  {125} {W} {WWWWWBW} {W}
\aff  {126} {W} {WWBBWWW} {W}
\aff  {127} {W} {BWWBWBW} {W}
\aff  {128} {W} {WWWBWWW} {W}
}\hskip 10pt
\vtop{\leftskip 0pt\parindent 0pt\hsize=\tabrulecol  
\ligne{\hfill\tt B $\rightarrow$ W\hfill}
\raff {129} {B} {WWBBBWW} {W}
\vskip 5pt
\ligne{\hfill\tt W $\rightarrow$ B\hfill}
\raff {130} {W} {BWWBWBB} {B}
\raff {131} {W} {WWBBBWW} {B}
\raff {132} {B} {BWWBWBW} {W}
}
\hfill}
\vskip 9pt
\trfn
\vskip 8pt
}
\hfill}

\ligne{\hfill
\vtop{\hsize=260pt
\begin{tab}\label{ectrlb}\leurre
Execution of the rules used during the traversal of a black controller by the locomotive.
\end{tab}
\vskip-2pt
\trep
\vskip 8pt
\ligne{\hfill\HH{}
\HH{{9$_6$} }\HH{{10$_6$}}\HH{{4$_6$} }\HH{{1$_6$} }\HH{{0$_0$} }\HH{{1$_4$} }\HH{{2$_5$} }\HH{{5$_5$} }\HH{{12$_4$}}\HH{{1$_3$} }
\hfill}
\ligne{\hfill\HH{1}
\HH{29}\HH{\Rr{17}}\HH{\Rr{63}}\HH{4}\HH{4}\HH{4}\HH{7}\HH{7}\HH{4}\HH{57}\hfill}
\ligne{\hfill\HH{2}
\HH{7}\HH{25}\HH{\Rr{31}}\HH{\Rr{45}}\HH{4}\HH{4}\HH{7}\HH{7}\HH{4}\HH{57}\hfill}
\ligne{\hfill\HH{3}
\HH{7}\HH{4}\HH{34}\HH{\Rr{17}}\HH{\Rr{36}}\HH{4}\HH{7}\HH{7}\HH{4}\HH{57}\hfill}
\ligne{\hfill\HH{4}
\HH{7}\HH{4}\HH{14}\HH{25}\HH{\Rr{17}}\HH{\Rr{36}}\HH{7}\HH{7}\HH{4}\HH{95}\hfill}
\ligne{\hfill\HH{5}
\HH{7}\HH{4}\HH{14}\HH{4}\HH{25}\HH{\Rr{17}}\HH{\Rr{16}}\HH{7}\HH{4}\HH{101}\hfill}
\ligne{\hfill\HH{6}
\HH{7}\HH{4}\HH{14}\HH{4}\HH{4}\HH{25}\HH{\Rr{24}}\HH{\Rr{16}}\HH{4}\HH{57}\hfill}
\ligne{\hfill\HH{7}
\HH{7}\HH{4}\HH{14}\HH{4}\HH{4}\HH{4}\HH{29}\HH{\Rr{24}}\HH{\Rr{36}}\HH{57}\hfill}
\vskip 9pt
\trfn
\vskip 8pt
}
\hfill}

\ligne{\hfill  
\vtop{\leftskip 0pt\parindent 0pt\hsize=340pt 
\begin{tab}\label{rctrlwit}\leurre
Rules for the cells~$0(0)$ and $1(3)$ of the controller. Rules for the passage of the
locomotive and for the signal, whatever the colour of the controller.
\end{tab}
\vskip-2pt 
\trep
\vskip 8pt
\ligne{\hskip 80pt passage\hfill signal\hskip 80pt}
\vskip 5pt
\ligne{0(0):\hfill   
\vtop{\leftskip 0pt\parindent 0pt\hsize=\tabrulecol
\ligne{\hfill black\hfill} %
\aff  {  4} {W} {WBWBWW\Gg B} {W}
\raff { 36} {W} {WB\Bb BBWW\Gg B} {B}
\raff { 17} {\Bb B} {WBWBWW\Gg B} {W}
\aff  { 25} {W} {\Bb BBWBWW\Gg B} {W}
\aff  {  4} {W} {WBWBWW\Gg B} {W}
}\hskip 15pt
\vtop{\leftskip 0pt\parindent 0pt\hsize=\tabrulecol
\ligne{\hfill white\hfill} %
\aff  { 77} {W} {WBWBWW\Rr W} {W}
\aff  {104} {W} {WB\Bb BBWW\Rr W} {W}
\aff  { 77} {W} {WBWBWW\Rr W} {W}
}\hskip 15pt
\vtop{\leftskip 0pt\parindent 0pt\hsize=\tabrulecol
\ligne{\hfill black\hfill} %
\aff  {  4} {W} {WBWBWW\Gg B} {W}
\aff  { 77} {W} {WBWBWW\Rr W} {W}
}\hskip 15pt
\vtop{\leftskip 0pt\parindent 0pt\hsize=\tabrulecol
\ligne{\hfill white\hfill} %
\aff  { 77} {W} {WBWBWW\Rr W} {W}
\aff  {  4} {W} {WBWBWW\Gg B} {W}
}\hfill}
\vskip 5pt
\ligne{1(3):\hfill   
\vtop{\leftskip 0pt\parindent 0pt\hsize=\tabrulecol
\ligne{\hfill black\hfill} %
\aff  { 57} {\Gg B} {WWBBWWW} {B}
\aff  { 95} {\Gg B} {\Bb BWBBWWW} {B}
\aff  {101} {\Gg B} {WWBBWW\Bb B} {B}
}\hskip 15pt
\vtop{\leftskip 0pt\parindent 0pt\hsize=\tabrulecol
\ligne{\hfill white\hfill} %
\aff  {126} {\Rr W} {WWBBWWW} {W}
}\hskip 15pt
\vtop{\leftskip 0pt\parindent 0pt\hsize=\tabrulecol
\ligne{\hfill black\hfill} %
\aff  { 57} {\Gg B} {WWBBWWW} {B}
\raff {129} {\Gg B} {WWBB\Oo BWW} {W}
\aff  {126} {\Rr W} {WWBBWWW} {W}
}\hskip 15pt
\vtop{\leftskip 0pt\parindent 0pt\hsize=\tabrulecol
\ligne{\hfill white\hfill} %
\aff  {126} {\Rr W} {WWBBWWW} {W}
\raff {131} {\Rr W} {WWBB\Oo BWW} {B}
\aff  { 57} {\Gg B} {WWBBWWW} {B}
}
\hfill}
\vskip 9pt
\trfn
\vskip 8pt
}
\hfill}

In Table~\ref{rctrlwit}, as in previous tables, black and white cells have different
meaning with respect to the simulation. For the convenience of the reader, we indicate
that \zz{\Bb B} marks the locomotive, \zz{\Oo B} marks the signal, \zz{\Gg B} shows us
that the controller is black which allows the passage of the locomotive while \zz{\Rr W}
shows us that it is white which forbids the passage of the locomotive. The way
the rules are working should now be clear without further comments.

Table~\ref{ectrlo} indicates the rules which are applied when the locomotive arrives
at a white controller and those which are applied when the signal for changing its colour
arrives at the controller.

\ligne{\hfill
\vtop{\leftskip 0pt\parindent 0pt\hsize=340pt
\begin{tab}\label{ectrlo}\leurre
Execution of the rules when the locomotive arrives to a white controller and
when the signal for changing the colour arrives.
\end{tab}
\vskip-2pt
\trep
\vskip 8pt
\ligne{\hfill
\hbox to 125pt{\hfill locomotive, 1(3) white\hfill}
\hskip 5pt \hbox to 180pt{\hfill signal for changing the colour\hfill}
\hfill}
\vskip 3pt
\ligne{\hfill
\vtop{\hsize=120pt
\ligne{\hfill\HH{}
\HH{{9$_6$} }\HH{{10$_6$}}\HH{{4$_6$} }\HH{{1$_6$} }\HH{{0$_0$} }
\hfill}
\ligne{\hfill\HH{1}
\HH{29}\HH{\Rr{17}}\HH{\Rr{63}}\HH{4}\HH{77}\hfill}
\ligne{\hfill\HH{2}
\HH{7}\HH{25}\HH{\Rr{31}}\HH{\Rr{45}}\HH{77}\hfill}
\ligne{\hfill\HH{3}
\HH{7}\HH{4}\HH{34}\HH{\Rr{17}}\HH{104}\hfill}
}
\hskip 5pt
\vtop{\hsize=90pt
\ligne{\hfill\tt W $\rightarrow$ B\hfill}
\vskip 3pt
\ligne{\hfill\HH{}
\HH{{12$_3$}}\HH{{4$_3$} }\HH{{1$_3$} }\HH{{1$_2$} }
\hfill}
\ligne{\hfill\HH{1}
\HH{25}\HH{\Rr{132}}\HH{\Rr{131}}\HH{125}\hfill}
\ligne{\hfill\HH{2}
\HH{4}\HH{78}\HH{57}\HH{121}\hfill}
}
\hskip 5pt
\vtop{\hsize=90pt
\ligne{\hfill\tt B $\rightarrow$ W\hfill}
\vskip 3pt
\ligne{\hfill\HH{}
\HH{{12$_3$}}\HH{{4$_3$} }\HH{{1$_3$} }\HH{{1$_2$} }
\hfill}
\ligne{\hfill\HH{1}
\HH{25}\HH{\Rr{92}}\HH{\Rr{129}}\HH{121}\hfill}
\ligne{\hfill\HH{2}
\HH{4}\HH{127}\HH{126}\HH{125}\hfill}
}
\hfill}
\vskip 9pt
\trfn
\vskip 8pt
}
\hfill}

\vtop{
\ligne{\hfill
\includegraphics[scale=0.55]{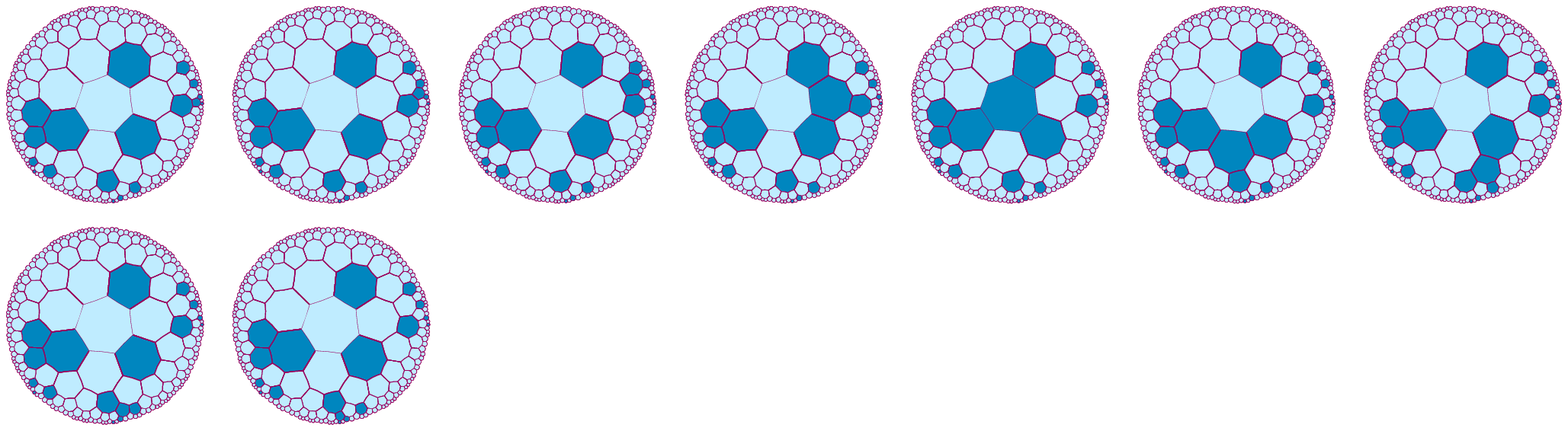}
\hfill}
\vspace{-70pt}
\ligne{\hskip 95pt
\includegraphics[scale=0.55]{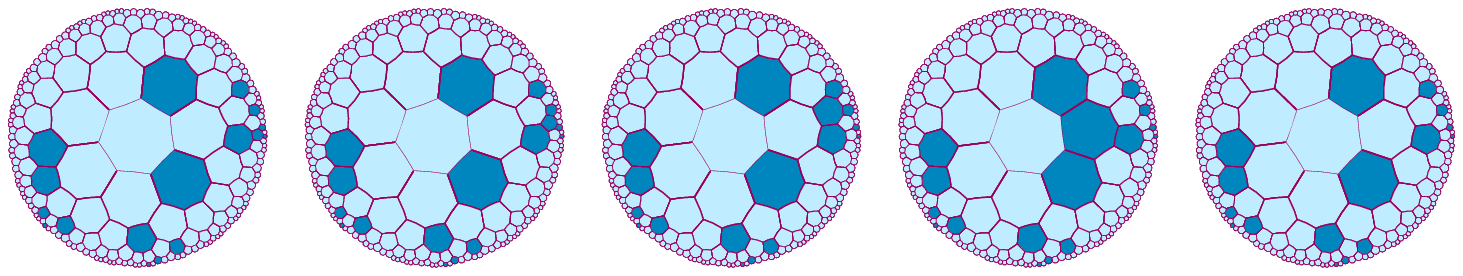}
\hfill}
\begin{fig}\label{fctrl}\leurre
Illustration of the crossing of the controller by the locomotive: above and first
figure of the second row, when it 
is black, below after the second figure, when it is white.
\end{fig}
}

\vtop{
\ligne{\hskip 70pt
\includegraphics[scale=0.55]{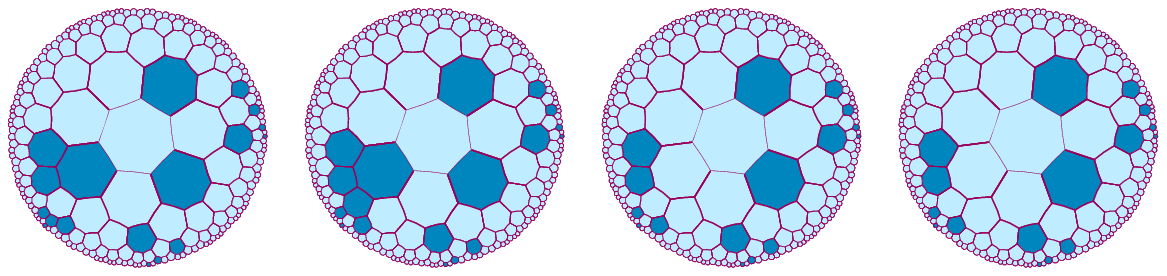}
\hfill}
\vspace{-20pt}
\ligne{\hskip 70pt
\includegraphics[scale=0.55]{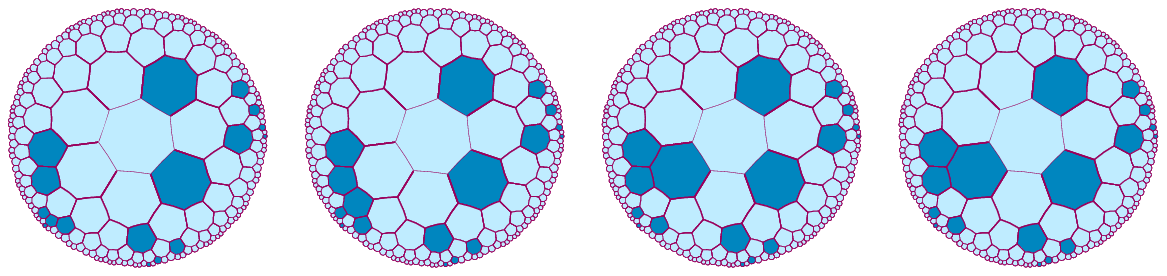}
\hfill}
\begin{fig}\label{fctrls}\leurre
Illustration of the arrival of the signal to the controller. Above: from black to white;
below: from white to black.
\end{fig}
}

Figure~\ref{fctrl} illustrates the crossing of controller by the locomotive
in both cases, according to the colour of~1(3). Figure~\ref{fctrls} illustrates the
arrival of the signal for changing the colour of the controller.

\subsection{The rules for the sensor}\label{srcapt}

In this last subsection of Section~\ref{rules}, we examine the rules which manage the
working of the sensor, the specific control structure of the passive memory switch.
Sub-section~\ref{sctrlcapt} in Section~\ref{scenar} explained the working of
the structure, pointing at the differences between the controller and the sensor
illustrated by Figures~\ref{stab_ctrl} and~\ref{stab_capt}.

Table~\ref{rcapt} illustrates the few rules which have to be appended to the already
examined 132 ones in order to make the structure working as expected.

\ligne{\hfill
\vtop{\leftskip 0pt\parindent 0pt\hsize=\tabrulecol
\advance\hsize by\tabrulecol\advance \hsize by \tabrulecol
\advance\hsize by 40pt  
\begin{tab}\label{rcapt}\leurre
Rules for the sensor of the passive memory switch.
\end{tab}
\vskip-2pt
\trep
\vskip 8pt
\ligne{\hfill \hbox to 120pt{\hfill passage\hfill}
\hbox to \tabrulecol{\hfill signal\hfill}\hfill}
\ligne{\hfill
\vtop{\leftskip 0pt\parindent 0pt\hsize=\tabrulecol  
\ligne{\hfill white\hfill}
\aff  {133} {W} {BWBWBWW} {W}
\aff  {134} {W} {WWWBBWW} {W}
\raff {135} {W} {BWBBBWW} {B}
\raff {136} {B} {BWBWBWW} {W}
\aff  {137} {B} {BWWWWBW} {B}
\aff  {138} {W} {BBBWBBW} {W}
}\hskip 10pt
\vtop{\leftskip 0pt\parindent 0pt\hsize=\tabrulecol  
\aff  {139} {W} {BWBWBBW} {W}
\aff  {140} {W} {BBWBBWB} {W}
\vskip 5pt
\ligne{\hfill black\hfill}
\aff  {141} {W} {BWBBBBW} {W}
\aff  {142} {B} {WBWWWBB} {B}
}\hskip 10pt
\vtop{\leftskip 0pt\parindent 0pt\hsize=\tabrulecol  
\ligne{\hfill\tt B $\rightarrow$ W\hfill}
\raff {143} {B} {WWWBBBW} {W}
\aff  {144} {B} {BBBWWWW} {B}
}
\hfill}
\vskip 9pt
\trfn
\vskip 8pt
}
\hfill}
\vskip 5pt

As can be seen in the comparison of Figures~\ref{stab_ctrl} and~\ref{stab_capt},
many rules used for the controller are also used for the sensor. As an example, as long
as the sensor is white, the rules executed in the cells of the tracks when the locomotive
passes are the same as those used in the same action when the controller is black,
see Tables~\ref{ectrlb} and~\ref{ecaptw}.

\ligne{\hfill
\vtop{\leftskip 0pt\parindent 0pt\hsize=240pt
\begin{tab}\label{ecaptw}\leurre
Execution of the rules when the sensor is white and then a locomotive passes.
\end{tab}
\vskip-2pt
\trep
\vskip 8pt
\ligne{\hfill\HH{}
\HH{{9$_6$} }\HH{{10$_6$}}\HH{{4$_6$} }\HH{{1$_6$} }\HH{{0$_0$} }\HH{{1$_4$} }\HH{{2$_5$} }\HH{{5$_5$} }\HH{{12$_4$}}\HH{{1$_1$} }
\hfill}
\ligne{\hfill\HH{1}
\HH{29}\HH{\Rr{17}}\HH{\Rr{63}}\HH{4}\HH{133}\HH{4}\HH{7}\HH{7}\HH{4}\HH{134}\hfill}
\ligne{\hfill\HH{2}
\HH{7}\HH{25}\HH{\Rr{31}}\HH{\Rr{45}}\HH{133}\HH{4}\HH{7}\HH{7}\HH{4}\HH{134}\hfill}
\ligne{\hfill\HH{3}
\HH{7}\HH{4}\HH{34}\HH{\Rr{17}}\HH{\Rr{135}}\HH{4}\HH{7}\HH{7}\HH{4}\HH{134}\hfill}
\ligne{\hfill\HH{4}
\HH{7}\HH{4}\HH{14}\HH{25}\HH{\Rr{136}}\HH{\Rr{36}}\HH{7}\HH{7}\HH{4}\HH{\Rr{131}}\hfill}
\ligne{\hfill\HH{5}
\HH{7}\HH{4}\HH{14}\HH{4}\HH{138}\HH{\Rr{17}}\HH{\Rr{16}}\HH{7}\HH{4}\HH{58}\hfill}
\ligne{\hfill\HH{6}
\HH{7}\HH{4}\HH{14}\HH{4}\HH{139}\HH{25}\HH{\Rr{24}}\HH{\Rr{16}}\HH{4}\HH{58}\hfill}
\ligne{\hfill\HH{7}
\HH{7}\HH{4}\HH{14}\HH{4}\HH{139}\HH{4}\HH{29}\HH{\Rr{24}}\HH{\Rr{36}}\HH{58}\hfill}
\vskip 9pt
\trfn
\vskip 8pt
}
\hfill}

However, as shown by Figures~\ref{stab_ctrl} and~\ref{stab_capt}, in the 
sensor, 1(3) is a milestone which, accordingly, is permanently black. 
The sensor cell is 1(1) which is either white or black, depending on the current
function of the sensor. We remind that the locomotive passes only if the sensor is white:
this means that the passing locomotive becomes a signal which in the sensor
provokes the change of the cell~1(1) from white to black and which will reach the other
sensor as a signal which will provoke the change of the other cell~1(1) from black
to white. When 1(1) is black and the locomotive arrives, it is stopped as
the current locomotive passed through the selected track arriving to the
passive memory switch: nothing has to be changed so that the duplicated locomotive has
not be changed into a signal to the other sensor.

Table~\ref{rcaptwit} shows the rules used for the cells 0(0) and 1(1). We use the same
conventions of colours as for the controller: as the global meaning of 
\zz B and of \zz W are the same, we keep the marks \zz{\Gg B} and \zz{\Rr W} for
the cell~1(1) which is the sensor cell of the structure.

\ligne{\hfill  
\vtop{\leftskip 0pt\parindent 0pt\hsize=340pt 
\begin{tab}\label{rcaptwit}\leurre
Rules for the cells~$0(0)$ and $1(3)$ of the controller. Rules for the passage of the
locomotive and for the signal, whatever the colour of the controller.
\end{tab}
\vskip-2pt 
\trep
\vskip 8pt
\ligne{\hskip 140pt passage\hfill signal\hskip 60pt}
\vskip 5pt
\ligne{0(0):\hfill   
\vtop{\leftskip 0pt\parindent 0pt\hsize=\tabrulecol
\ligne{\hfill white\hfill} %
\aff  {133} {W} {BWBWB\Rr WW} {W}
\raff {135} {W} {BWB\Bb BB\Rr WW} {B}
\raff {136} {\Bb B} {BWBWB\Rr WW} {W}
\aff  {138} {W} {B\Bb BBWB\Gg BW} {W}
\aff  {139} {W} {BWBWB\Gg BW} {W}
}\hskip 15pt
\vtop{\leftskip 0pt\parindent 0pt\hsize=\tabrulecol
\ligne{\hfill black\hfill} %
\aff  {139} {W} {BWBWB\Gg BW} {W}
\aff  {141} {W} {BWB\Bb BB\Gg BW} {W}
\aff  {139} {W} {BWBWB\Gg BW} {W}
}\hskip 15pt
\vtop{\leftskip 0pt\parindent 0pt\hsize=\tabrulecol
\ligne{\hfill black\hfill} %
\aff  {139} {W} {BWBWB\Bb BW} {W}
\aff  {133} {W} {BWBWB\Rr WW} {W}
}\hfill}
\vskip 5pt
\ligne{1(1):\hfill   
\vtop{\leftskip 0pt\parindent 0pt\hsize=\tabrulecol
\ligne{\hfill white\hfill} %
\aff  {134} {\Rr W} {WWWBBWW} {W}
\raff {131} {\Rr W} {WW\Bb BBBWW} {B}
\aff  { 58} {\Gg B} {WWWBBWW} {B}
}\hskip 15pt
\vtop{\leftskip 0pt\parindent 0pt\hsize=\tabrulecol
\ligne{\hfill black\hfill} %
\aff  { 58} {\Gg B} {WWWBBWW} {B}
}\hskip 15pt
\vtop{\leftskip 0pt\parindent 0pt\hsize=\tabrulecol
\ligne{\hfill black\hfill} %
\aff  { 58} {\Gg B} {WWWBBWW} {B}
\raff {143} {\Gg B} {WWWBB\Oo BW} {W}
\aff  {134} {\Rr W} {WWWBBWW} {W}
}
\hfill}
\vskip 9pt
\trfn
\vskip 8pt
}
\hfill}

\ligne{\hfill
\vtop{\leftskip 0pt\parindent 0pt\hsize=250pt
\begin{tab}\label{ecaptb}\leurre
Execution of the rules for the black sensor, for the locomotive and for the signal.
\end{tab}
\vskip-2pt
\trep
\vskip 8pt
\ligne{\hfill
\vtop{\leftskip 0pt\parindent 0pt\hsize=130pt
\ligne{\hfill locomotive\hfill}
\ligne{\hfill\HH{}
\HH{{9$_6$} }\HH{{10$_6$}}\HH{{4$_6$} }\HH{{1$_6$} }\HH{{0$_0$} }
\hfill}
\ligne{\hfill\HH{1}
\HH{29}\HH{\Rr{17}}\HH{\Rr{63}}\HH{4}\HH{139}\hfill}
\ligne{\hfill\HH{2}
\HH{7}\HH{25}\HH{\Rr{31}}\HH{\Rr{45}}\HH{139}\hfill}
\ligne{\hfill\HH{3}
\HH{7}\HH{4}\HH{34}\HH{\Rr{17}}\HH{141}\hfill}
\ligne{\hfill\HH{4}
\HH{7}\HH{4}\HH{14}\HH{4}\HH{139}\hfill}
}
\hskip 10pt
\vtop{\leftskip 0pt\parindent 0pt\hsize=110pt
\ligne{\hfill signal\hfill}
\ligne{\hfill\HH{}
\HH{{9$_1$} }\HH{{3$_1$} }\HH{{1$_1$} }\HH{{2$_1$} }
\hfill}
\ligne{\hfill\HH{1}
\HH{29}\HH{\Rr{92}}\HH{\Rr{143}}\HH{144}\hfill}
\ligne{\hfill\HH{2}
\HH{7}\HH{127}\HH{134}\HH{15}\hfill}
\ligne{\hfill\HH{3}
\HH{7}\HH{127}\HH{134}\HH{15}\hfill}
}
\hfill}
\vskip 9pt
\trfn
\vskip 8pt
}
\hfill}

Table~\ref{rcaptwit} shows us several features. First, due to the choice of side~1
shared with 1(3), the neighbourhood of 0(0) is now \hbox{\tt 1, 3, 5}, see rule~133
in Tables~\ref{rcapt} and~\ref{rcaptwit}. However, after the applications of rules~135
and~136 working as front and cell rules, the next rule is~138 as the passage of the
locomotive in~0(0) triggered the change of~1(1) from \zz{\Rr W} to 
\zz{\Gg B}, see rule~138
in Table~\ref{rcaptwit} followed by rule~139, the conservative rule of~0(0)
when the controller is black. This can be seen in the column devoted for~0(0) to
the passage of the locomotive in a black controller.

A second feature deals with the cell~1(1): its neighbourhood is \hbox{\tt 4, 5}. The rule
is conservative when the sensor is black, see rule~58, it is not when it is white~:
see rule~131. Cells with only two milestones and contiguous ones were already met in 
the motion on the tracks as rules~18 and~21 or rules~121 and~126 for the controller.
In all those latter rules which are conservative the current state of the cell is white.
This again point at the importance of the relaxation of the rotation invariance 
hypothesis.

The third feature is another fact about the rules for~1(1) in Table~\ref{rcaptwit}: the
change of coloured is not triggered in the same way. From white to black, the change
is triggered by the locomotive, see rule~131 where the black cell of the locomotive
is marked with \zz{\Bb B}. From black to white, it is triggered by the 
signal, see rule~143 where the locomotive-signal is marked \zz{\Oo B}.

\vtop{
\ligne{\hfill
\includegraphics[scale=0.55]{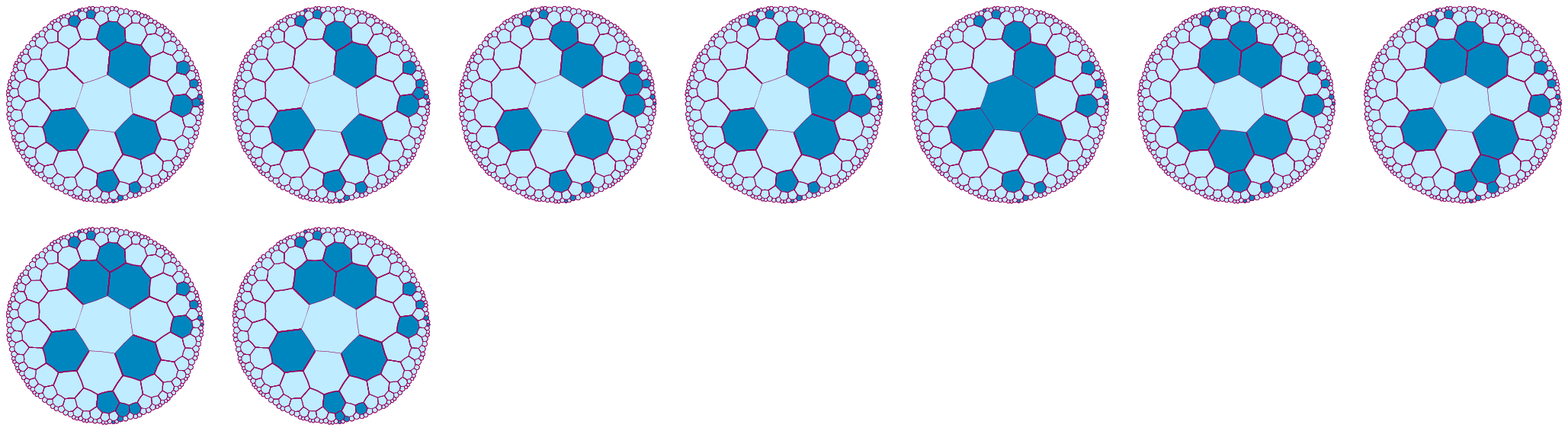}
\hfill}
\vspace{-20pt}
\ligne{\hfill
\includegraphics[scale=0.55]{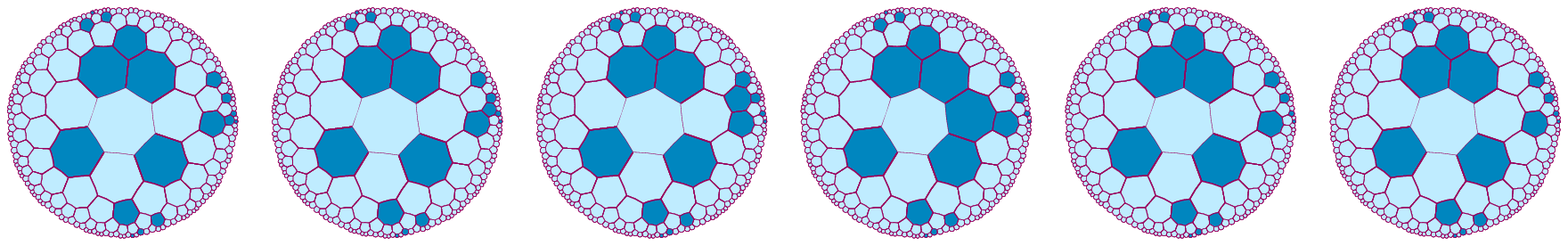}
\hfill}
\vspace{-20pt}
\ligne{\hfill
\includegraphics[scale=0.55]{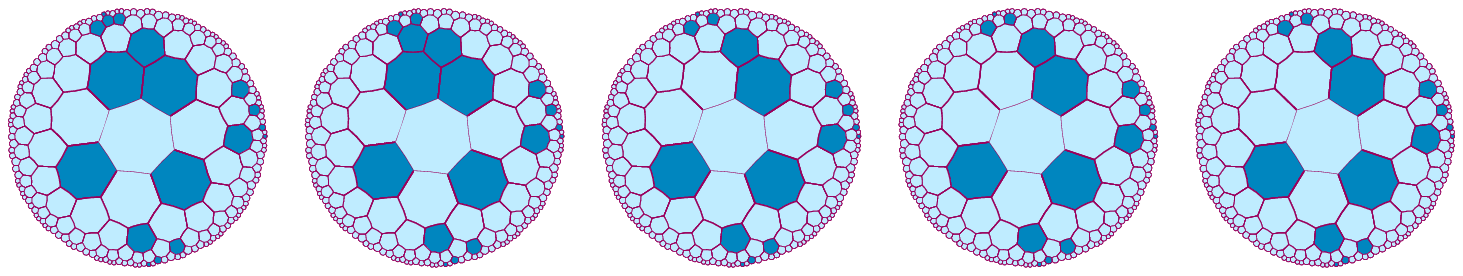}
\hfill}
\begin{fig}\label{fcapt}\leurre
Illustration of the working of the sensor. First row: passage of the locomotive;
second row: the locomotive is stopped; third row: the signal changes a black sensor
to a white one.
\end{fig}
}

We completed the examination of the rules for the sensor.
Accordingly, Theorem~\ref{letheo} is proved.
\hfill\boxempty

\section*{Conclusion}

As mentioned in the introduction, Theorem~\ref{letheo} is the best result for 
tessellations involving regular convex polygon with the angle 
$\displaystyle{{2\pi}\over3}$: the tessellation $\{7,3\}$ is the tessellation $\{p,3\}$ 
where $p$ has the smallest value as possible for the hyperbolic plane. With this model, 
the implementation in the heptagrid seems to be impossible under the rotation invariance 
assumption. There are two rules with no black neighbour and also two ones
with exactly one black neighbour. The neighbourhoods are:

\ligne{\hfill\zz{W$^7$} and  \zz{BW$^6$}.\hfill} 
\vskip 3pt\noindent
There are also tow rules with no white neighbour and two ones with one white neighbour
exactly: they are obtained by \textbf{contraposition} from the just mentioned ones,
\textit{i.e.} by exchanging \zz B and \zz W.
There are six rules with either two black neighbours exactly, which are

\ligne{\hfill \zz{B$^2$W$^5$}, \zz{BWBW$^4$} and \zz{BW$^2$BW$^3$}.\hfill}
\vskip 3pt\noindent
The neighbours with two white neighbours exactly are obtained by contraposition.
There are ten rules three black neighbours exactly:

\ligne{\hfill \zz{B$^3$W$^4$}, \zz{B$^2$WBW$^3$}, \zz{B$^2$W$^2$BW$^2$}, 
\zz{B$^2$W$^3$BW} and \zz{BWBWBW$^2$}.\hfill}
\vskip 5pt\noindent 
By contraposition we obtain the neighbourhoods with three white neighbours exactly. 
Accordingly, there at most forty rotation independent rules. 
There are already big problems to define the tracks as our solution with three
milestones produces a single rotation invariant rule and, as noted in 
Subsection~\ref{srtrack}, several rules are rotationally incompatible. With 
four-milestoned rules, we have a problem with symmetric patterns which cannot
be used for one way motions. For the same reason of symmetry, we cannot use a
pattern with five milestones as the two white neighbours for entry and exit could not
be distinguished.

\end{document}